\documentclass[a4paper,11pt]{article}

\pdfoutput=1
\usepackage{jheppub}
\usepackage[T1]{fontenc}
\usepackage[british]{babel}
\usepackage{rotating}
\usepackage{ifpdf}
\usepackage{booktabs}

\title{\boldmath Higgs Pair Production: Choosing Benchmarks With Cluster Analysis}

\author[a]{Alexandra Carvalho, }
\author[a]{Martino Dall'Osso, }
\author[a]{Tommaso Dorigo, }
\author[b]{Florian Goertz,}
\author[c]{Carlo A. Gottardo, }
\author[b]{Mia Tosi }

\affiliation[a]{Dipartimento di Fisica e Astronomia and INFN, Sezione di Padova, Via Marzolo 8, I-35131 Padova, Italy}
\affiliation[b]{CERN, 1211 Geneva 23, Switzerland}
\affiliation[c]{Physikalisches Institut, Universit{\"a}t Bonn, Nussallee 12, 53115 Bonn, Germany}

\abstract{
   New physics theories often depend on a large number of free parameters. The phenomenology they predict for fundamental physics processes is in some cases drastically affected by the precise value of those free parameters, while in other cases is left basically invariant at the level of detail experimentally accessible. When designing a strategy for  the analysis of experimental data in the search for a signal predicted by a new physics model, it appears advantageous to categorize the parameter space describing the model according to the corresponding kinematical features of the final state. A multi-dimensional test statistic can be used to gauge the degree of similarity in the kinematics predicted by different models; a clustering algorithm using that metric may allow the division of the space into homogeneous regions, each of which can be successfully represented by a benchmark point. Searches targeting those benchmarks are then guaranteed to be sensitive to a large area of the parameter space.

   In this document we show a practical implementation of the above strategy for the study of non-resonant production of Higgs boson pairs in the context of extensions of the standard model with anomalous couplings of the Higgs bosons. A non-standard value of those couplings may significantly enhance the Higgs boson pair-production cross section, such that the process could be detectable with the data that the LHC will collect in Run 2.
}
\usepackage[
            style=numeric-comp, 
            firstinits=true, 
            maxcitenames=300,
            maxbibnames=500,
            block=none,
            sortcites=true,  
            sorting=none,
            backend=bibtex
            ]{biblatex}
            
\begin{document} 
\maketitle
\flushbottom

\section{Introduction}
\vskip .3cm

After the Run 1 discovery of a new scalar particle at 125 GeV~\cite{higgsdiscoveryAtlas,Chatrchyan:2012ufa} the LHC experiments are now looking forward to the data they are collecting in Run 2 and in the higher-luminosity phases that will follow. New discoveries are possible with the significantly increased centre-of-mass energy of proton-proton ({\em pp}) collisions and the foreseen integrated luminosity. The new data will also enable a deep investigation of the 125 GeV particle. To test if the latter can be identified with the Higgs boson predicted by the Standard Model (SM)~\footnote{ In this article we follow the terminology which has become standard in high-energy physics, namely we call ``Higgs boson'' the scalar particle resulting from the Brout-Englert-Higgs (BEH) mechanism when adding a complex doublet of scalar fields to the unbroken SM Lagrangian. }, and very generally to probe the mechanism of electroweak symmetry breaking (EWSB), it is of the utmost importance to measure the scalar potential. 

In the SM Lagrangian the Higgs potential contains a quartic self interaction of the Higgs doublet. The interplay of this term with the negative-sign mass term $-\mu^2$ drives electroweak symmetry breaking (EWSB) (see however~\cite{Goertz:2015dba}). One physical state, the Higgs boson {\em h}, remains in the theory, with cubic and quartic self-couplings resulting from the quartic interaction of the scalar doublet. The measurement of these self-couplings, possible with the study of multi-Higgs production, allows us to gain information on the scalar potential. In the SM scenario the strength of all Higgs boson couplings is precisely predicted; deviations from those predictions would thus imply the existence of beyond-Standard-Model (BSM) physics. A high-statistics study of the couplings of the newly discovered boson may therefore reveal whether we are in the presence of the last building block of the SM, or rather of the first one of a new physics sector.

The idea of probing BSM physics scenarios (especially the Higgs trilinear coupling) in non-resonant Higgs boson pair production at proton-proton colliders dates far before the top quark discovery and the LHC design~\cite{Glover:1987nx}; a large number of studies have been performed since then. After the Higgs boson discovery, many authors have investigated different phenomenological aspects of the topic (see for example~\cite{Dawson:2015oha,Azatov:2015oxa,  Slawinska:2014vpa, Barger:2013jfa, Barr:2014sga, Liu:2014rba, Goertz:2014qta, Chen:2014xra, Dolan:2012ac, Goertz:2013kp,  Nishiwaki:2013cma, Yao:2013ika,Shao:2013bz,Wardrope:2014kya,Dolan:2015zja,He:2015spf,Dolan:2015zja,Lu:2015jza, Cao:2013si,Cao:2014kya,Papaefstathiou:2015iba,deLima:2014dta,Maierhofer:2013sha,Papaefstathiou:2012qe}); most of the phenomenology-driven works have focused on the effects of a variation of the Higgs boson trilinear coupling $\lambda$.  In the present work we consider that any kind of coupling deviation from the SM Higgs sector is a proof that the SM is not complete; therefore {\em all} possible Higgs boson couplings should be considered as ingredients of a BSM extension of the Standard Model~\footnote{It is interesting to note that by measuring Higgs boson pair production, one can even probe the very presence of the $\mu^2$ term, {\em i.e.} the relevant $H^2$ operator (before electroweak symmetry breaking)~\cite{Goertz:2015dba}.}. The parameter space resulting from that interpretation is multi-dimensional; its systematic study calls for a principled approach, which we aim to provide here. 

In this paper we will focus on gluon-gluon fusion (GF) production of Higgs boson pairs, which is the simplest process available to probe the Higgs boson self-coupling at the LHC. The possible BSM deviations in inclusive di-Higgs production arising in other production modes, such as vector-boson fusion~\cite{Hankele:2006ma,VBFhh} or associated production with top quarks (see for example~\cite{Englert:2014uqa,Liu:2014rva}) or vector bosons, probe a different set of parameters in the context of an effective-field theory (EFT) as the one we are considering. The interpretation of the results of those channels should be studied separately. 

Measurements performed by the ATLAS and CMS collaborations with Run 1 LHC data have already started to constrain the value of some of the Higgs boson couplings~\cite{Khachatryan:2014jba, ATLAS-CONF-2015-007}. Due to interference effects, even small modifications of some of the couplings within the constraints posed by measurements may change drastically the di-Higgs signal topology, and enhance the production cross section enough to make the process accessible with Run 2 data. For that to happen, the observable features of the final state need to be exploited in an optimized way, given the huge cross section of physical processes yielding irreducible backgrounds. This implies making full use of the distinguishing characteristics of signal events in the multi-dimensional space of their observable features. What is needed is therefore the identification of a manageably small set of benchmark points which are maximally representative of the largest possible volume of the unexplored parameter space. The investigation of those points in as much detail as possible will effectively provide information on the full model space. 

We employ a very general parametrization to sample the di-Higgs signal topology, assuming the absence of new heavy particles accessible at the LHC energy, which can for example describe the effects of strongly-coupled BSM theories (without being limited to those); we provide it in Sec.~\ref{sec:lag}. The rest of this work is organized as follows. In Sec.~\ref{sec:cluster} we describe in detail the technique we devised to determine the similarity of final state densities in the space of observable kinematics, and the clustering procedure which uses that measure of similarity to identify homogeneous regions in the parameter space. In Sec.~\ref{sec:res} we describe the application of the technique to determine optimal benchmarks for the study of anomalous di-Higgs boson production, and we discuss the special features of the resulting partition of the parameter space. We finally draw some conclusions in Sec.~\ref{sec:conclu}. In the Appendix we also provide the coefficients of our parametrization of the di-Higgs production cross section.

\section{Sampling Signal Kinematics in the Higgs Couplings Basis}
\label{sec:lag}
\vskip .3cm

In the SM Higgs boson pair production occurs predominantly by gluon-gluon fusion (GF) via an internal fermion loop, where the top quark contribution is dominant. This is because the Higgs boson couplings are exclusively controlled by the particle masses; couplings to light quarks are negligible. 
The extension of the latter feature as an assumption for BSM theories is well motivated if the Higgs sector is minimal (see also~\cite{Goertz:2014qia}).  
In the absence of new light states, the GF Higgs boson pair production at the LHC can then be generally described (to leading approximation) by five parameters controlling the tree-level interactions of the Higgs boson. These five parameters, which will be discussed in detail in the following, are $\kappa_{\lambda}$, $\kappa_{t}$, $c_g$, $c_{2g}$, and $c_2$. The Higgs boson trilinear coupling and the top Yukawa interaction do exist in the SM Lagrangian, where the former is given by $\lambda_{SM}=m_h^2/2v^2$, with $v$ the vacuum-expectation value of the Higgs field. Deviations from SM values are parametrized with the multiplicative factors $\kappa_{\lambda}$ and $\kappa_{t}$, respectively. The contact interactions of the Higgs boson with gluons and those coupling two Higgs bosons with two gluons or a top-antitop quark pair, which could arise through the mediation of very heavy new states, are instead genuinely not predicted by the SM; they can be parametrized by the absolute couplings $c_g$, $c_{2g}$, and $c_2$. The relevant part of the Lagrangian then takes the form \par

\begin{eqnarray}
{\cal L}_h = 
\frac{1}{2} \partial_{\mu}\, h \partial^{\mu} h - \frac{1}{2} m_h^2 h^2 -
  {\kappa_{\lambda}}\,  \lambda_{SM} v\, h^3 
- \frac{ m_t}{v}(v+   {\kappa_t} \,   h  +  \frac{c_{2}}{v}   \, h\,  h ) \,( \bar{t_L}t_R + h.c.) \nonumber  \\ 
+ \frac{1}{4} \frac{\alpha_s}{3 \pi v} (   c_g \, h -  \frac{c_{2g}}{2 v} \, h\, h ) \,  G^{\mu \nu}G_{\mu\nu}\,.
\label{eq:lag}
\end{eqnarray}

\begin{figure}[h]
\centering
\includegraphics[scale=0.85]{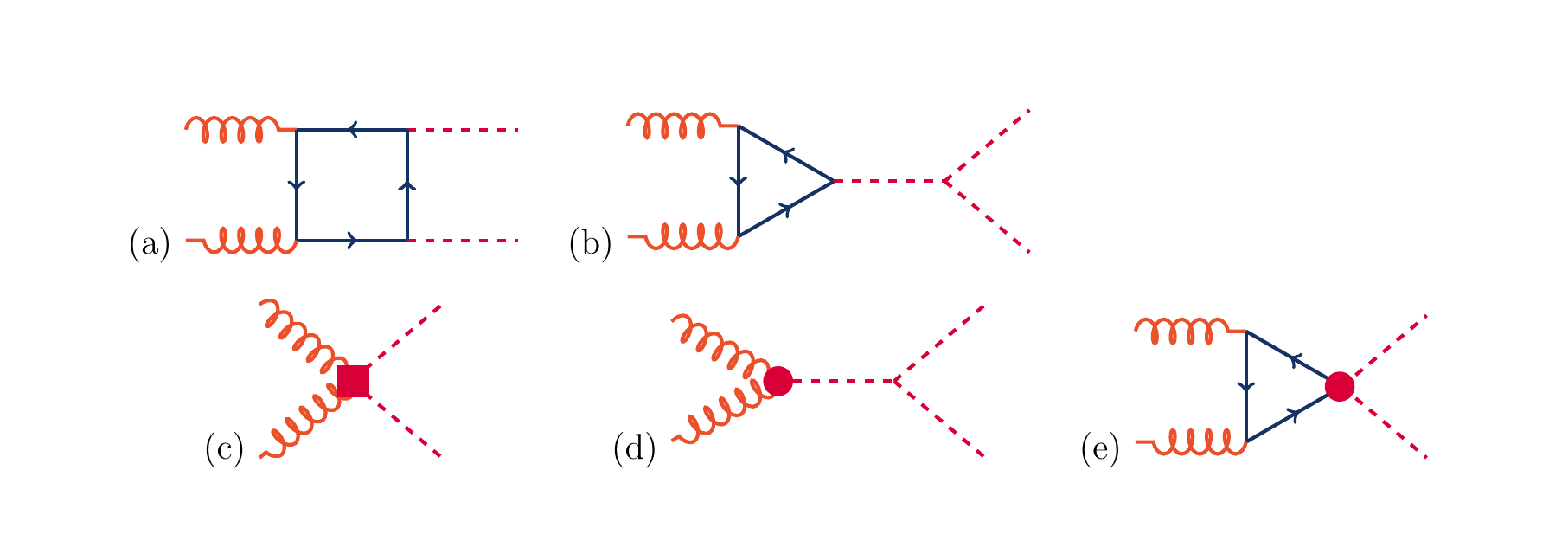}
\caption{\small Feynman diagrams that contribute to Higgs boson pair production by gluon-gluon fusion at leading order. Diagrams (a) and (b) correspond to SM-like processes, while diagrams (c), (d), and (e) correspond to pure BSM effects: (c) and (d) describe contact interactions between the Higgs boson and gluons, and (e) exploits the contact interaction of two Higgs bosons with top quarks.   \label{fig:dia}}
\end{figure}

\noindent
In fact, this Lagrangian follows from extending the SM with operators of mass dimension $4<D\leq6$ in the framework of an effective field theory (EFT), encoding the effects of new heavy states currently beyond experimental reach. In the case of a linear realization of EWSB, one obtains the EFT relation $c_{2g} = - c_g$~\cite{Giudice:2007fh,Buchmuller:1985jz,Contino:2013kra,Gillioz:2012se}~\footnote{Our normalization for the Higgs boson interaction with gluons is inspired by the infinite top-mass limit of the SM. The existence of a relative sign between $c_{2g}$ and  $c_g$ in this limit is a special feature of the SM, related to the chiral nature of SM fermions. }.
In Eq.~\ref{eq:lag} we have assumed the absence of any other light state in addition to the SM particles. In the presence of such states, the kinematic structures will in general be further modified~\cite{Dawson:2015oha}\footnote{In models with an extended (non-decoupled) Higgs sector, the coupling of the Higgs boson to bottom quarks might also be strongly enhanced in the limit of large scalar mixing. The topology of double Higgs boson production would consequently be modified: besides the presence of a component of the signal initiated by bottom fusion, the gluon-fusion topology would be modified since loop factors would now contain a non-negligible component with a low-mass quark~\cite{Plehn:1996wb}. An enhanced Higgs boson coupling to bottom quarks is not the only physical effect that could arise in Supersymmetric (SUSY) theories, where additional SUSY scalars are predicted (see for example~\cite{Belyaev:1999kk,Spira:1997ce,Han:2013sga}). In the context of SUSY scenarios non-resonant di-Higgs boson production  is a wide topic, and we do not discuss it further in this document, although our treatment may still be useful in the decoupling limit, where all new states are heavy.}. In addition, we do not consider CP-violating BSM effects. Finally, we recall that the bottom quark Yukawa coupling $\kappa_b$ in the EFT is already constrained within $0.75<\kappa_b<1.25$ by LHC data~\cite{Corbett:2015ksa}, excluding large enhancements and justifying its omission in our framework as a good approximation. The different Feynman diagrams contributing to a di-Higgs boson signal in {\em pp} collisions at leading order (LO) are shown in Fig.~\ref{fig:dia}.

\noindent
In Eq.~\ref{eq:lag} we included operators with higher orders of the Higgs boson fields, which are for example a common feature of models where the Higgs doublet is a pseudo-Goldstone boson of a new strong (broken) symmetry and its effective interactions come from field expansions~\cite{Dugan:1984hq,Giudice:2007fh}. The translation of our parametrization to the flavour-diagonal Higgs basis (see~\cite{Falkowski:2014tna, Falkowski:2015fla}), which has been endorsed by the LHCHXSWG document~\cite{Duehrssen-Debling:2001958} as a general EFT basis to be used to derive experimental results, is trivial; for simplicity we prefer to keep the notation of Eq.~\ref{eq:lag}. Any dimension-6 EFT basis is related to the Higgs basis by analytical relations among the coefficients; the automation of basis conversions is under development~\cite{roseta}.  

The differential cross section of the full process under consideration is proportional to the matrix element squared. We may write the square of the full matrix element (ME) at LO as\par

\begin{eqnarray}
| M_{full}|^2 = 
 |M_{\lambda}|^2 + |M_{\Box}|^2 +  |M_{c_2}|^2 +  |M_{c_g}|^2 +  |M_{c_{2g}}|^2 + \nonumber \\   
 \, (M_{\lambda} M_{\Box}^{\dagger})  + 
  \, (M_{\lambda} M_{c_2}^{\dagger})  + 
   \, (M_{\lambda} M_{c_g}^{\dagger})  +  
    \, (M_{\lambda} M_{c_{2g}}^{\dagger}) + \nonumber \\ 
  \, (M_{\Box} M_{c_2}^{\dagger})  + 
   \, (M_{\Box} M_{c_g}^{\dagger})  + 
    \, (M_{\Box} M_{c_{2g}}^{\dagger})  + \nonumber \\
   \, (M_{c_2} M_{c_g}^{\dagger})  + 
    \, (M_{c_2} M_{c_{2g}}^{\dagger})  + 
    \, (M_{c_g} M_{c_{2g}}^{\dagger})  + \rm{ h.c}. \,.
    \label{eq:ME}
\end{eqnarray}

\noindent
where the various terms in the matrix element expression contribute differently in different regions of the kinematic space. Above, $M_{j}$ identifies the matrix element piece where the parameter $j$ is entering at LO, and $M_{\Box}$ corresponds to the box diagram. Ideally, the bulk of the higher-order corrections do not spoil this structure to reasonable approximation.

\subsection{Cross Section \label{sec:CX1}}

The full cross section of GF Higgs boson pair production can be expressed by a polynomial in terms of all the model parameters as\par 

\begin{eqnarray}
\frac{\sigma_{hh}}{\sigma_{hh}^{SM}} =   \left( 
\begin{matrix}
A_1\, \kappa_t ^4 + A_2\, c_2^2 + (A_3\, \kappa_t^2 + A_4\, c_g^2)\, \kappa_{\lambda}^2 
+ A_5\, c_{2g}^2 + ( A_6\, c_2 + A_7\, \kappa_t \kappa_{\lambda} )\kappa_t^2  \\
+ (A_8\, \kappa_t \kappa_{\lambda} + A_9\, c_g \kappa_{\lambda} ) c_{2} + A_{10}\, c_2 c_{2g} 
+ (A_{11}\, c_g \kappa_{\lambda} + A_{12}\, c_{2g})\, \kappa_t^2  \\
+ (A_{13}\, \kappa_{\lambda} c_g + A_{14}\, c_{2g} )\, \kappa_t \kappa_{\lambda} + A_{15}\, c_{g} c_{2g} \kappa_{\lambda} 
\end{matrix} 
\right)
\label{eq:cx}
\end{eqnarray}

\noindent
In proton-proton collisions at 13 TeV the SM prediction is $\sigma_{hh}^{SM} = 34.3 \mbox{ fb  }\pm 9\% \mbox{ (scale)} \pm 2\%$(PDF), while at 8 TeV it is $\sigma_{hh}^{SM} = 9.96 \mbox{ fb  } \pm 10\% \mbox{ (scale)}  \pm 2\%$ (PDF)\footnote{The cross sections have been very recently calculated at NNLL in QCD~\cite{YR4,deFlorian:2015moa,deFlorian:2013jea}.}. Those values are based on recent studies~\cite{deFlorian:2014rta,Grigo:2014jma} which use the {\tt CT10} PDF set~\cite{Gao:2013xoa} and employ as input the mass values $m_h = 126$ GeV, $m_t = 173.18$ GeV, and $m_b = 4.75$ GeV. 
At LO the scattering amplitude for the $gg\to hh$ process contains terms with different loop structures, corresponding to the different operators. The real emissions for the $gg\to hh$ process are not trivial to compute; the corresponding diagrams would contain up to pentagons to be matched with parton showers. Different groups of phenomenologists are progressing in the calculation of (N)NLO predictions matched to shower-level effects for the GF di-Higgs boson production process, especially for the SM case; see for example~\cite{Maltoni:2014eza,Frederix:2014hta,Grigo:2015dia,deFlorian:2015moa,Grober:2015cwa}. 

The simulation setup used in this paper was produced by the authors of~\cite{Hespel:2014sla}. The LO process is already at one-loop level; in the approach followed in~\cite{Hespel:2014sla}, loop factors are calculated on an event-by-event basis with a {\tt Fortran} routine on top of an $aMC@NLO$~\cite{Frixione:2010ra, Alwall:2014hca} effective model; the {\tt NN23LO1} PDF set~\cite{Ball:2009mk} is used. Those simulations represent the state-of-the art in the description of BSM di-Higgs boson production. We simplify the task of mapping the five-dimensional parameter space of BSM theories in the limit of the present computational capability by assuming that each of the matrix element pieces of Eq.~\ref{eq:ME} gets corrected by an overall k-factor, as written in the first equality of Eq.~\ref{eq:NNLO}. As a second step we make the stronger assumption that the k-factors related to the different ME pieces are equal, and that they may be taken as a k-factor derived for the SM case, leading to the second equality in Eq.~\ref{eq:NNLO}:

\begin{equation}
 (M_{i} M_{j}^{\dagger} + h.c.)^{\mbox{\small higher order}} = k_{ij} \, (M_{i} M_{j}^{\dagger} + h.c.)^{(LO)}
 = k_{SM} \, (M_{i} M_{j}^{\dagger} + h.c.)^{(LO)}\,.
 \label{eq:NNLO}
\end{equation}

\noindent
The above approximations are expected to be good for QCD-like radiative corrections when quoting the total cross section. Indeed, the enhancement in the total cross section from QCD NLO corrections is mainly due to soft gluon radiation from the initial state~\cite{Grober:2015cwa}. For a characterization of the differential distributions, on the other hand, the description outlined above might not be entirely satisfactory. Bearing in mind that potential caveat, we decided to use it for this study as it facilitates the mapping of experimental results derived with LO simulations to the results of a radiative corrected calculation.

\begin{figure}[h]
\centering
\includegraphics[scale=0.75]{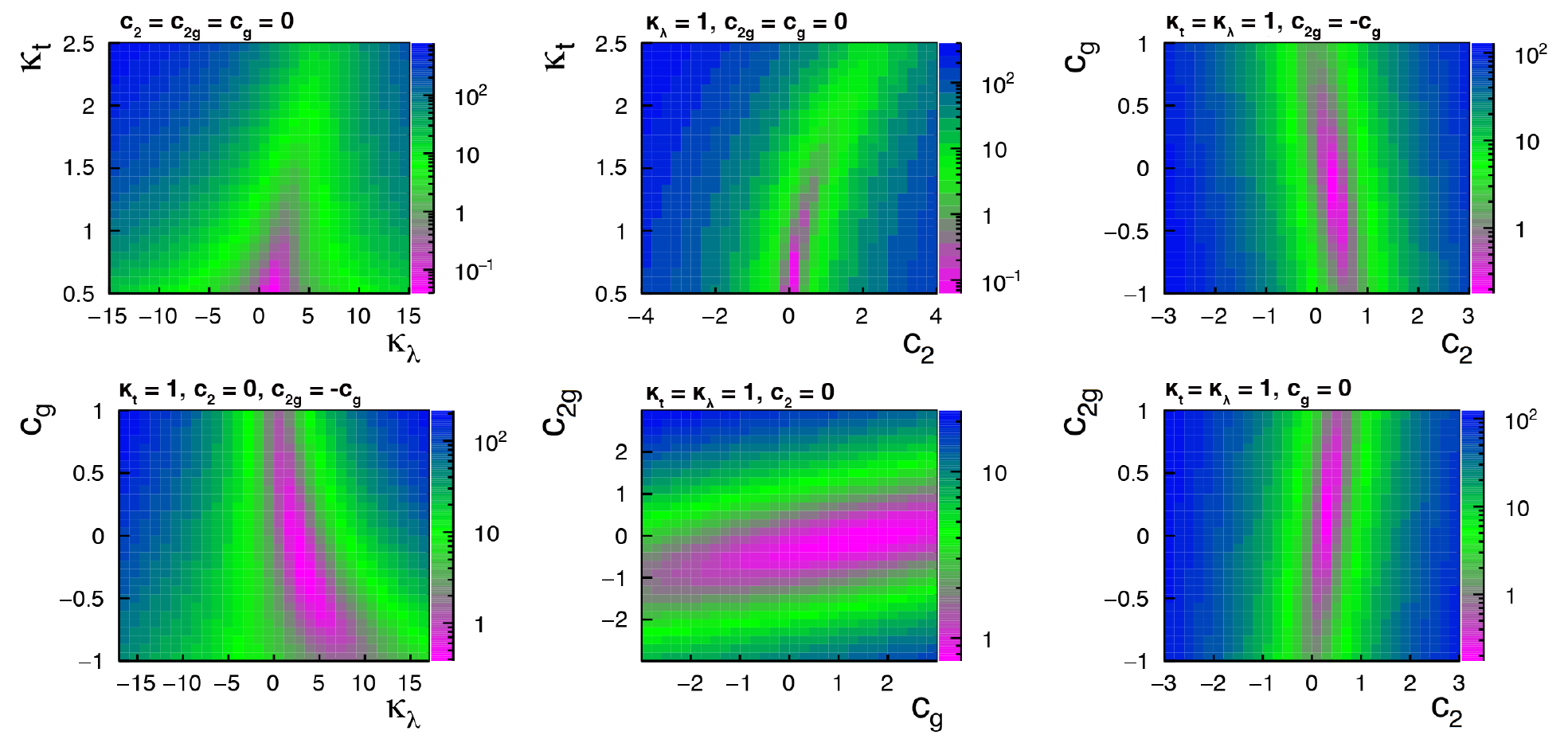}
\caption{\small  Cross section ratios ($\sigma_{BSM}/\sigma_{SM}$) in selected slices of parameter space. Left column: the plane of SM parameters, $\kappa_t : \kappa_\lambda$ (top), and the region allowing a Higgs boson contact interaction with gluons, $c_g : \kappa_\lambda$ (bottom). Middle column: planes spanned by the parameters describing non-vanishing one- and two-Higgs boson interactions with top quarks and with gluons, $\kappa_t : c_2$ (top) and $c_{2g} : c_g$ (bottom). Right column: the planes spanned by parameters governing interactions of the Higgs boson with gluons and top-quark pairs, $c_g : c_2$ (top) and $c_{2g}:c_2$ (bottom), for selected values of the other parameters. The cross section is computed with the fit discussed in the text.
\label{fig:cx}
}
\end{figure}

Using Eq.~\ref{eq:NNLO} it is possible to calculate the coefficients of the polynomial~\ref{eq:cx} by  evaluating the results of LO computations in different points of the five-dimensional parameter space. For each considered point, using the setup mentioned above, we generate 20,000 {\em pp} collision events at 13 TeV centre of mass energy, producing a final state of two Higgs bosons. The resulting cross sections are then fit with a maximum likelihood technique to the polynomial~\ref{eq:cx}. In order to ensure a stable fit we inspect six orthogonal two-dimensional planes in the five-dimensional parameter space that all contain the point corresponding to the SM. The procedure used to derive the coefficients of the polynomial and the numerical results for the fitted parameters is detailed in~\cite{CarvalhoAntunesDeOliveira:2130724,}. Figure~\ref{fig:cx} shows the resulting cross section in the two-dimensional planes mentioned above.
The range of parameters considered in our study is discussed below.

\subsection{Parameter Space Study \label{sec:CX2}}

In order to carry out a phenomenological study of Higgs boson pair production in BSM theories we need to first define the range of parameter variations we are willing to consider. Some of the parameters relevant to the production phase are basically unconstrained by single Higgs boson measurements: among these are the triple Higgs coupling and the di-Higgs boson contact interactions with top quarks and gluons. Others, such as the top Yukawa coupling, are already constrained by experimental results~\cite{Khachatryan:2014jba, ATLAS-CONF-2015-007}. A precise interpretation of all experimental bounds as constraints on the effective operators and their effect on the considered process is not trivial, as the parameters do not only affect the predicted rate of di-Higgs boson production, but also the kinematics of the final state. 

Measurements of single Higgs boson production performed so far at the LHC already constrain the $\kappa_{t}$ and $c_g$ parameters. The combination of those results using the $\kappa$ formalism~\cite{Artoisenet:2013puc} shows that, by marginalising over all other Higgs boson couplings, the allowed values of $\kappa_{t}$ are constrained at $95\%$ C.L. in the region between 0.5 and 2.5.  A Bayesian analysis of BSM operators based on available measurements of Higgs boson properties constrains the $c_g$ parameter to be at most at the $O(1)$ level~\cite{Dumont:2013wma,Elias-Miro:2013mua,Pomarol:2013zra,Gupta:2014rxa,Corbett:2015ksa}. The remaining parameters are constrained only by absolute cross section limits on inclusive di-Higgs boson production~\cite{Aad:2014yja} as explained below.

The current experimental limits on non-resonant Higgs boson pair production in the SM come from the study of the $\gamma\gamma\, b\bar{b}$ and $4b$ final states at 8 TeV by ATLAS~\cite{Aad:2014yja,Aad:2015uka}. Those limits are respectively $\sigma_{hh\to\gamma\gamma bb}^{SM} < 5.72$ fb (220 times the SM value) and $\sigma_{hh\to 4b}^{SM} < 202$ fb (56 times the SM value). Both results above were derived by counting experiments and they involve no strong assumptions on the signal topology in the final state other than the presence of two Higgs bosons. Considering the ATLAS results extended to all the parameter space, we find the $|\kappa_{\lambda}|$-only variation to be constrained at 95\% C.L. in the region $|\kappa_{\lambda}| \lesssim 15$~\footnote{The $\kappa_{\lambda}$ parameter describes a multiplicative variation of a small value ($\lambda_{SM} \sim 0.13$), therefore an $O(10)$ variation would not affect the computational validity of the perturbative approach. Beyond that, note that the theoretical range of validity as an EFT is to some extent model-dependent. While in a weakly coupled scenario large BSM contributions to the coefficients signal a low new-physics scale, setting the energy cutoff of the theory, and thus limit the applicability of the setup, in a strongly coupled scenario sizable coefficients can arise consistently with a high cutoff. } . Following a similar approach  the $c_2$ parameter can be constrained to $|c_2|<5$ at $95\%$ CL when $\kappa_{\lambda} = 1$ and $\kappa_{t} \subset [0.5\, , 2.5]$. 


A cursory look at the kinematics of the final state, as described in any of the already cited phenomenological studies, suggests that different choices of the coupling parameters give rise to striking differences in the density functions of the kinematic observables. This convinces us of the need of a systematic approach to characterize the signal topology.

In order to retain generality of the results of our study for any final state of di-Higgs boson production, and invariance to further analysis cuts and/or analysis techniques, we study the event topology as it results from the production of the two Higgs bosons free from initial-state radiation effects and before the subsequent decays and final-state radiation effects. The study is performed with an extended sampling of parameter space points with respect to the ones used to calculate the total cross section in last section; again, no generation cut is applied to the processes. For each studied point of the parameter space we generate 20,000 events of {\em pp} collisions at 13 TeV centre of mass energy.  These are sufficient for the task of understanding how the event kinematics varies as a function of the model parameters.

We are considering a $2\to 2$ process at leading order. The two Higgs bosons are produced with identical transverse momenta ($p_{T}^{h}$), and they are back-to-back in azimuth at this order (before a parton shower). The final state can then be completely defined by three kinematic variables, if we ignore the irrelevant azimuthal angle of emission of the bosons. Furthermore, one of the three remaining variables can be used to isolate all the information related to the PDF of the colliding partons, which is also irrelevant to the physics of the production process once one focuses on a specific initial state (the gluon-gluon fusion process). The variable factorizing out the PDF modelling can be taken as the magnitude of the boost of the centre of mass frame as seen in the laboratory frame. 

The two remaining variables, which provide direct information on the physics of GF di-Higgs boson production, can be chosen to be the invariant mass of the di-Higgs system ($m_{hh}$) and the modulus of the cosine of the  polar angle of one Higgs boson with respect to the beam axis ($|cos\theta^*|$). Since we are using parton-level information, this last variable is equivalent to the polar angle in the Collins-Soper frame ($|cos\theta^*_{CS}|$)~\cite{PhysRevD.16.2219}, which is commonly used in experimental analysis. The variables $m_{hh}$ and $|cos\theta^*|$ can thus be used to fully characterize the final state kinematics produced by different choices of the value of anomalous Higgs boson (self-) coupling parameters.

\section{Classification of Final State Kinematics} \label{sec:cluster}
\vskip .3cm

The choice of benchmarks for the study of a new physics model is usually a difficult task,  as it obliges to several partly conflicting desires. While the collection of benchmarks should in principle offer an exhaustive representation of the varied final state composition and topologies that the new physics model may give rise to, one's choice of the specific values of the model parameters to study in more detail often falls on those which are within the sensitivity reach of a specific amount of data collected by a given experiment at a given time. In that case the focus is usually on the cross section of the new physics signal, which is identified as the most important factor. As it happens with the drunkard who lost his watch in the dark and only searches it under the street lamps, this approach is guaranteed the highest short-term impact but may not be the most principled if one takes a long-term perspective.

The case of Higgs boson pair production at the LHC offers a peculiar situation, as in the short term we will be unable to achieve experimental sensitivity to the largest part of the BSM parameter space. Furthermore, anomalous Higgs boson pair production processes are characterized by a final state which is homogeneous in its composition, as opposed to, {\em e.g.}, SUSY production processes, which give rise to a quite rich and diverse set of final states depending on the exact choice of theory parameters. Within that homogeneous final state, anomalous di-Higgs boson production offers quite varied kinematics as a function of the model parameters. This makes it an ideal ground for a principled and quantitative approach to the choice of benchmark points. 

In light of the above considerations we take the problem from the side of shape information rather than normalization. By identifying sets of parameters which yield similar final state kinematics we simplify the problem of investigating a large and unconstrained model space. The resulting partition of the space will remain useful as the integrated luminosity collected by the LHC experiments grows from tens to hundreds of inverse femtobarns. 

The task of partitioning the parameter space into homogeneous regions can be performed with cluster analysis techniques. These allow the grouping of elements of a set into subsets in such a way that members of each subset are mutually more similar to one another than are elements belonging to different subsets. The similarity will, in our case, be described by an ordering parameter which is constructed with the event kinematics. 

\subsection{Two-Sample Tests} 
\vskip .3cm

In order to define a metric to classify physics models based on the similarity of the event kinematics they describe in the feature space, we need to choose a general statistical framework as well as a suitable two-sample test statistic. At first, we might consider the problem as one of hypothesis testing. Accordingly, we would define a test size $\alpha$ and a null hypothesis $H_0^{ij}$ for each pair of parameter space points $i$ and $j$, the null hypothesis being that the corresponding data samples $S_i$ and $S_j$ share the same parent probability density function --or, in other words, that models $i$ and $j$ describe the same physics. The choice of a test statistic and its evaluation on all pairs of samples would then allow us to populate a matrix describing the mutual compatibility of the samples, in the form of a set of pass/fail bits. Clearly, such a result would not be practical for the task of grouping samples into subsets of similar characteristics. Furthermore, it must be noted that as we start with samples which do originate from distinguishable parameter space points and which yield different density functions, it is only the lack of infinite statistics what might prevent us from calling two samples passing the test as ``different'' from an experimental point of view. 

We may turn the limited statistics of the datasets to our advantage if we realize that what we need is an analog answer rather than a digital one: a degree of similarity between each pair of samples must take the place of the yes/no answer of the hypothesis test. A test statistic (TS) such as a $\chi^2$ probability or a likelihood value may be used to determine which samples should be grouped into homogeneous subsets. 

There exists a large variety of two-sample tests suitable for the task at hand. To name a few, one may use the Anderson-Darling test~\cite{ADart}, the Kolmogorov-Smirnov test, the $\chi^2$ test, the T test, or others. The ones mentioned above are usually single-dimensional tests, in the sense that they are meant to compare two single-dimensional distributions; their extension to multi-dimensional data is not always straightforward, as it is subject to implementation choices that call for detailed power studies~\footnote {The power of a test, $1-\beta$, is the probability that the test is capable of evidencing the truth of the alternative hypothesis, as $\beta$ is the type-2 error rate, {\em i.e.} the probability that the test rejects the alternative hypothesis when in fact it is the true one.}. In a multi-dimensional setup possible choices also include the Energy test~\cite{Aslan:2002cn} or nearest-neighbour-based metrics. Such unbinned multi-dimensional TS may be the right choice in situations when the statistics of the samples to be compared are very small, or when the dimensionality of the problem is large. In the specific case of non-resonant di-Higgs boson production, however, we found that the TS with highest power to detect localized differences in the kinematic distributions is a likelihood ratio based on Poisson counts in a set of two-dimensional bins. That is the solution we investigate and discuss in this work. 

\subsection{The Likelihood Ratio Test Statistic} {\label{s:TS} 

\vskip .3cm

In the specific application described here, the numerousness of our generated datasets (20,000 events per sample) and the small dimensionality of the feature space that completely defines the final state of the process (two variables) allow us to employ as test statistic a binned likelihood ratio.
The number of bins in $m_{hh}$ and $|\cos\theta^*|$ are chosen such that the main kinematic features of the distributions are properly modelled while retaining sufficiently populated bins. We found appropriate for our application to have fifty 30-GeV-wide bins in $m_{hh}$ in the range from zero to 1500 GeV and five 0.2-wide bins in $|\cos\theta^*|$ from zero to one.

To define our test statistic let us first consider the hypothetical case in which the two samples under test share the same parent distribution. The corresponding likelihood function is the product over the bins of the probability to observe $n_{i,1}$ and $n_{i,2}$ event counts in bin $i$ from the two samples $S_1$ and $S_2$. This probability is given by the product of two Poisson distributions $Pois(n_{i,1} | \hat{\mu_{i}}) \times Pois(n_{i,2} | \hat{\mu_{i}})$ where $\hat{\mu_{i}} =  (n_{i,1}+n_{i,2})/2$ is the maximum likelihood estimate for the expected contents in bin $i$. However it can be shown that

\begin{equation}
Pois(n_{i,1}) \times Pois(n_{i,2}) = Pois(n_{i,1}+n_{i,2}) \times Binomial(n_{i,1}/(n_{i,1}+n_{i,2})).
\end{equation}

\noindent
It is clear that the first term in the right-hand side of the decomposition does not contain any information about the differences between the density functions of the two samples; it only contains information on the precision of the test. This is what is called, in statistics literature, an \emph{ancillary statistic} which can be advantageously neglected, as we do in the following. The retained binomial term is explicitly\par

\begin{equation}
Binomial(n_{i,1}/( n_{i,1} + n_{i,2})) = \frac{(n_{i,1}+n_{i,2})!}{n_{i,1}!n_{i,2}!}\left(\frac{1}{2}\right)^{n_{i,1}} \left(\frac{1}{2}\right)^{n_{i,2}}.
\label{eq:P0}
\end{equation}

\noindent
Now, to obtain a likelihood ratio we consider the case in which the two samples are equal, the so-called \emph{saturated hypothesis}~\cite{bakercousins}. The appropriate single-bin-content probability can be obtained from Eq.~\ref{eq:P0} by imposing $n_{i,1}=n_{i,2}=\hat{\mu_{i}}$, yielding\par

\begin{equation}
Binomial(n_{i,1}=n_{i,2}=\hat{\mu_{i}}) = \frac{(2\hat{\mu_{i}})!}{(\hat{\mu_{i}}!)^{2}}\left( \frac{1}{2}\right)^{2\hat{\mu_{i}}}.
\label{eq:Psat}
\end{equation}

\noindent
Calling $L$ the likelihood obtained from the distribution in Eq.~\ref{eq:P0} and $L_{S}$ the one from Eq. \ref{eq:Psat} we define the log-likelihood ratio\par

\begin{equation}
TS = 2 \ log\left(\frac{L}{L_{S}} \right) = -2 \sum_{i=1}^{N_{bins}} \left[ log(n_{i,1}!) + log(n_{i,2}!) -2log\left(  \frac{n_{i,1}+n_{i,2}}{2}! \right) \right],
\label{eq:TS}
\end{equation}

\noindent
which, up to a minus sign, is "$\chi^{2}$ distributed"~\cite{bakercousins,Wilks:1938dza}. Thanks to this property this TS can be directly used as an ordering parameter to perform a cluster analysis. In other words, the values $TS_{ij}$ and $TS_{kl}$ obtained respectively by testing the compatibility of samples $ij$ and $kl$ are suitable to determine if samples $S_i$ and $S_j$ are more similar to each other than are  samples $S_k$ and $S_l$: this is the case if $TS_{ij}>TS_{kl}$~\footnote{For a generic test statistic which is not distribution-independent, this is not granted; one is then forced to study the probability density function of the TS under the null hypothesis  for each pair of tested distributions, comparing $p$-values derived from tail integrals of the TS. Besides being extremely CPU consuming, this also requires to use part of the data to construct the null distribution of the TS for each sample pair. }. 

In addition to its distribution-independence the TS of Eq.~\ref{eq:TS} is particularly sensitive to small-scale features of the distributions under test, and is thus well suited to our task as we are confronted with samples exhibiting bi-modal structures in the studied spectra (see for instance Fig.~\ref{fig:mlj}). In contrast, TS which are more sensitive to large-scale structure may give precedence to it when used as an ordering parameter in a clustering procedure: we have observed that such behaviour gives rise to unwanted results, whereby bimodal and single-modal distributions are clustered together. 

\subsection{The Clustering Technique}
\label{sec:clu}
\vskip .3cm
The clustering procedure must produce a grouping of the parameter space points based on the kinematical distributions of the corresponding final states. Such a task can be performed in a number of ways, yielding in general different results. The algorithm we chose matches our desire to create homogeneous regions of parameter space based on the TS metric, and it allows to univocally identify the sample in each cluster which is the most representative of the set - what we call a {\it benchmark}. The benchmark is chosen as the sample which is the most similar to all the other samples associated to the same cluster.

The sample comparisons are pairwise, therefore from $N_{sample}$ tested points we can form $N_{sample}(N_{sample}-1)/2$ two-sample test results with the procedure described in Sec.~\ref{s:TS}. We define the following procedure to group samples into a given number of clusters ($N_{clus}$):\par

\begin{enumerate}
\item Start by identifying each of the $N_{sample}$ elements as one-element clusters.
\item Define the cluster-to-cluster {\it similarity} as $TS^{min} = min_{ij} (TS_{ij})$, where $i$ runs on all elements of the first cluster and $j$ runs on all
elements of the second cluster. 
\item Find among all the possible pairs of clusters the pair with the highest value of $TS^{min}$; merge the two clusters into one, and recompute the resulting benchmark (see below).
\item Repeat step 3 above until $N_{clus}$ clusters are left, keeping a record of all intermediate results.
\item Identify the benchmark sample in a cluster as the element $k$ with the highest value of $TS_{k}^{min} = min_{i} (TS_{ki})$ between the clustered samples, where $i$ runs on all elements of the cluster except $k$ (if more elements have the same value of $TS^{min}_{k}$, one may by convention take the first one).
\end{enumerate}

\noindent
Figure~\ref{f:clustering} describes graphically the clustering method. For any given choice of the number of clusters the procedure returns the optimal clustering  and the benchmark in each cluster. Of course, there is a trade-off between intra-cluster homogeneity and $N_{clus}$: as the latter decreases, more and more discrepant elements are clustered together; accordingly, the benchmark becomes less and less representative on the whole of the subset that contains it.

It is easy to see how the technique outlined above possesses some attractive features for our application. There is always a well-defined benchmark in each cluster, and the criterion by which points are clustered together privileges a maximum intra-cluster uniformity over an average one. In the next section we apply the method to the parameter space points describing BSM di-Higgs boson production, which allows us to show what those properties mean in practice.

\begin{figure}[h!]
\centering
\includegraphics[scale=0.65]{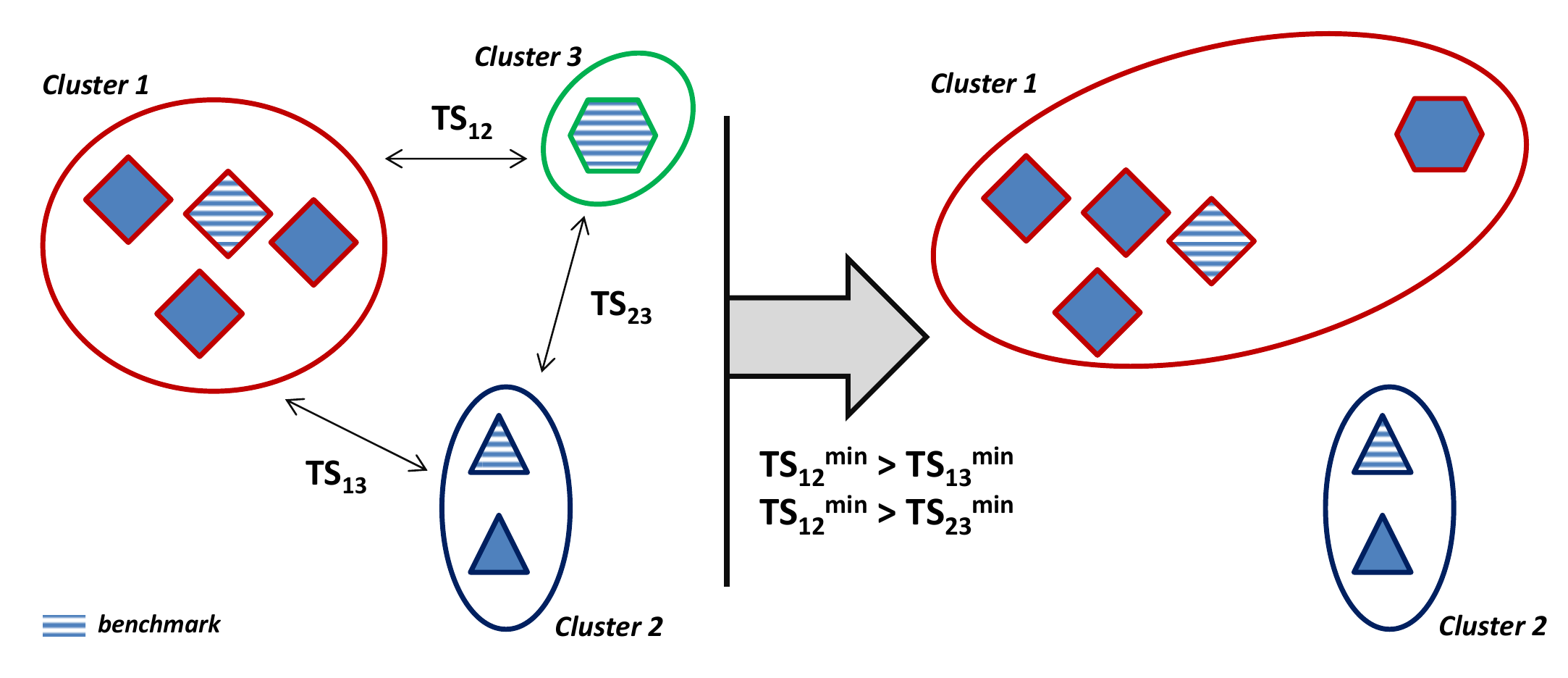}
\caption {\small Graphical description of the clustering procedure.}
\label{f:clustering}
\end{figure}

\section{Application to Higgs Pair Production}
\label{sec:res}
\vskip .3cm

In this section we discuss the application of the procedure described in Sec.~\ref{sec:cluster} to GF di-Higgs boson production at the LHC. The first step is to identify the set of parameter space points on which we wish to run the cluster analysis. Ideally one would like to start with a regular and homogeneous grid in the five-dimensional parameter space of anomalous couplings described in Sec.~\ref{sec:lag}; however any meaningfully-spaced regular grid would require a prohibitive number of simulated data samples. 
Instead of using a regularly spaced grid, we focus primarily on the regions of parameter space where the probability densities of the final state observables exhibit the fastest variability with parameter variation. These regions coincide with local minima of the production cross section, as explained below (Sec.~\ref{sec:temp}). The resulting population of the five-dimensional grid is admittedly arbitrary; however it is seen {\em a posteriori} to be able to picture reasonably well the varied spectrum of topologies of GF di-Higgs boson production. It includes $N_{sample}=1507$ points of the five-dimensional parameter space, composed of the following three subsets:

\begin{itemize}
   \item We start with a geometrically well-spaced grid in the slices of Fig.~\ref{fig:cx}, identified by values of $\kappa_{\lambda} = 0,\,  \pm 1,\, \pm 2.4,\, \pm 3.5,\, \pm 5 ,\, \pm 10,\, \pm 15$; $\kappa_{t}$ from $0.5$ to $2.5$ in steps of $0.25$ when $|\kappa_{\lambda}| < 5$, and steps of $0.5$ elsewhere; $c_2$ between $-3.0$ and $3.0$ in steps of 0.5; $c_g$ and $c_{2g}$ between $-1.0$ and $1.0$ in steps of 0.2. 
   \item In some regions of parameter space (especially those with $c_2 =0.5$ and $c_{2g}={\cal O}(1)$) there is a strong cancellation between the different operators in the threshold $m_{hh}$ region. This leads to topologies where the distribution of $m_{hh}$ exhibits a long tail to high values~\footnote{We thank A.~Papaefstathiou for checking the same kinematic behaviour using an alternative model implementation in Herwig.}. In order to have a better kinematic description of this topology (and as well of the cancellation pattern between operators) we add to the grid one slice of parameter space with $c_2 = 0.5$ and $\kappa_{\lambda}=\kappa_{t} = 1$, maintaining the previous binning in the $c_g - c_{2g}$ plane.  
   \item Finally, we also consider a three-dimensional grid of points described by the parameters $\kappa_{\lambda}$, $\kappa_{t}$, and $c_2$ in the hyperplane defined by $c_g = c_{2g} = 0$. The points are identified by combinations of the following parameter values: $\kappa_{\lambda} = \pm 1,\, \pm 2.4,\,\pm 3.5,\, \pm 5 ,\,\pm 7.5,\,$ $\pm 10,\,\pm 12.5,\, \pm 15$; $\kappa_{t}$ from $0.5$ to $2.5$ in steps of $0.25$; and $c_2$ between $-3.0$ and $3.0$ in steps of $0.5$. An increased density of points is allocated near the point corresponding to the SM hypothesis ($|c_2|<1$). 
\end{itemize}

\noindent
Figures~\ref{fig:tempSM}, \ref{fig:temp3D}, and \ref{fig:5Dgridcg} in Sec.~\ref{sec:temp} show graphically the location of the generated parameter space points.
 
\subsection{Choice of the Number of Clusters}

The total number of required clusters $(N_{clus})$, and therefore the total number of regions into which the parameter space is divided, is the only free parameter in the clustering procedure described in Sec.~\ref{sec:cluster}. The uniformity of the kinematical distributions within each cluster is a qualitative criterion which can be used to choose the target value of $N_{clus}$. A large number of clusters provides a fine sub-division of the parameter space and guarantees a better uniformity of the kinematical distributions within each cluster. However, a too large number of benchmarks puts a heavy load on the experimental treatment of the data needed to probe the full parameter space. On the other hand, a too small $N_{clus}$ may produce marked differences in the samples grouped together, such that the corresponding benchmark does not appear suitable to accurately represent the behaviour of the subset. 

In our specific application we have observed that strong discrepancies within the clusters appear when $N_{clus}$ becomes smaller than $12$, while for $N_{clus}>12$ the differences between the kinematical distributions of the samples included in different clusters are small enough that they should have a limited impact on the extrapolation of results obtained for the benchmark point. Figure~\ref{fig:evol} shows the $m_{hh}$ distribution for the two clusters that are merged when the number of clusters is reduced by one unity from $N_{clus}=13$ to 9. It is evident that when reducing $N_{clus}$ from 13 to 12 there is no significant worsening of the uniformity of the merged cluster, while the same cannot be said in further reducing $N_{clus}$. Given the good uniformity of the distributions in all clusters, $N_{clus} = 12$ is the value chosen for the cluster analysis of the 1507 samples of di-Higgs boson production model points. We consider this a reasonable trade-off between homogeneity and numerousness of the clusters.

\begin{figure*}[h]\begin{center}
\begin{tabular}{l l}
\includegraphics[width=0.48\textwidth, angle =0 ]{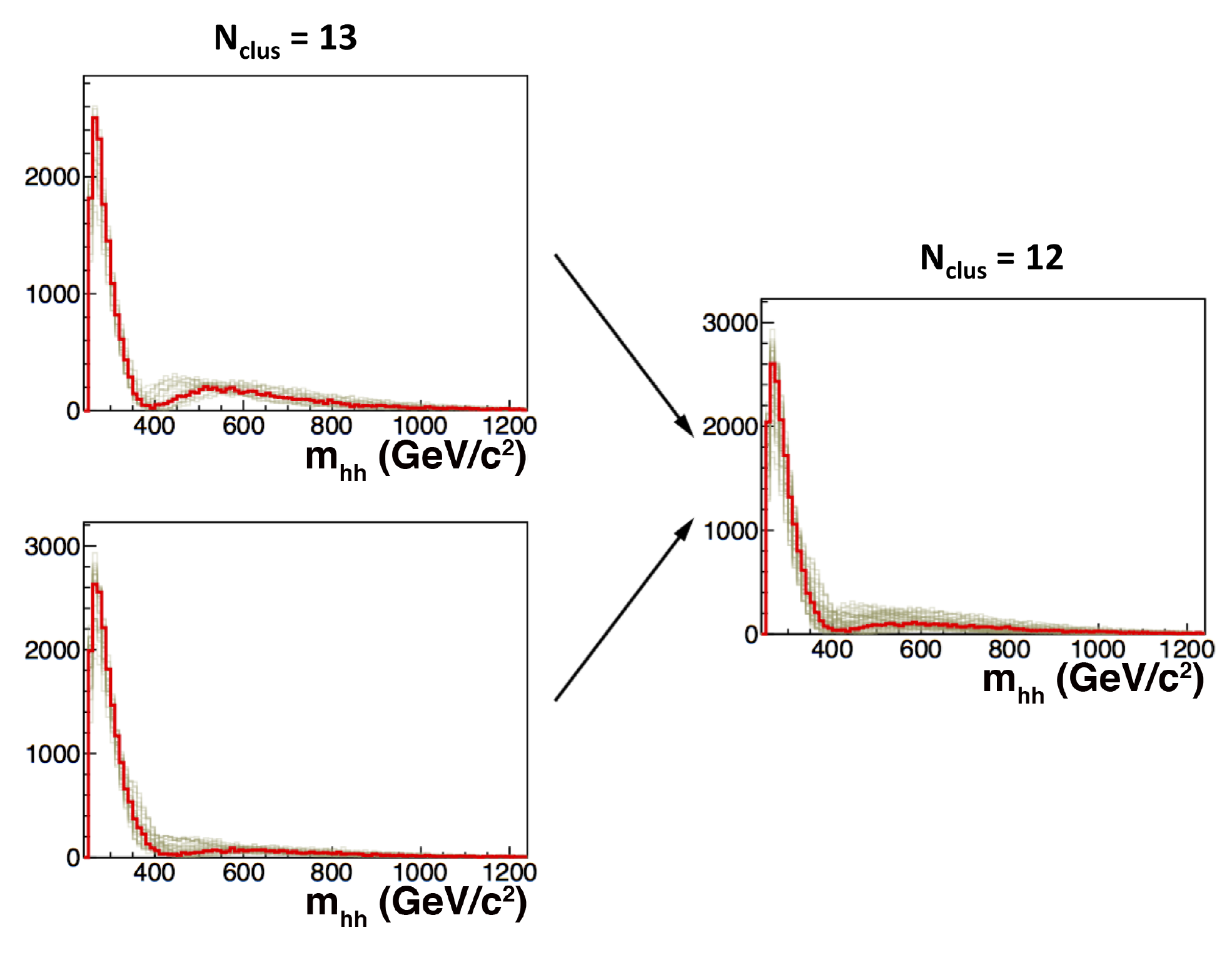}
 &
 \includegraphics[width=0.48\textwidth, angle =0 ]{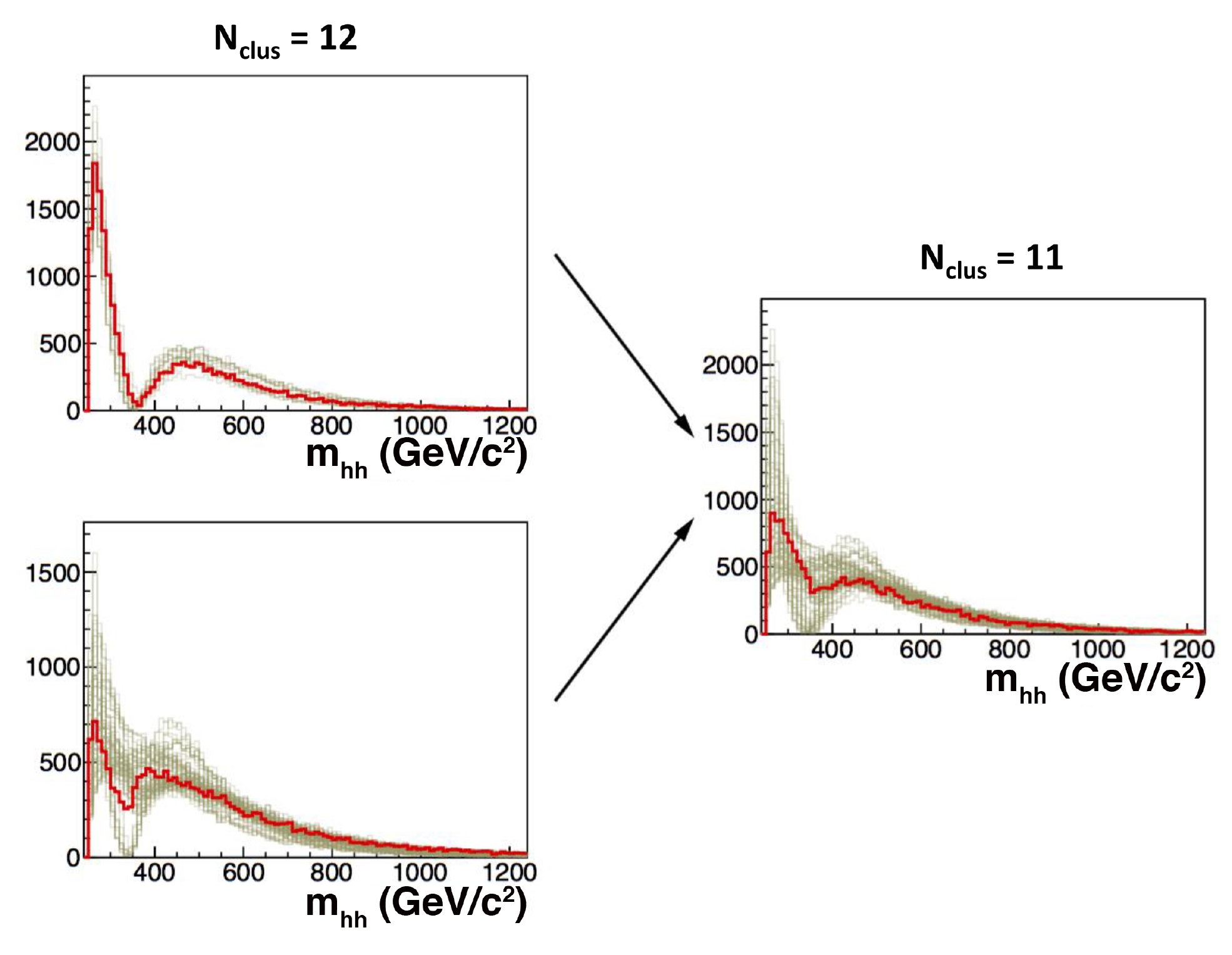} \\
\hline
 \includegraphics[width=0.48\textwidth, angle =0 ]{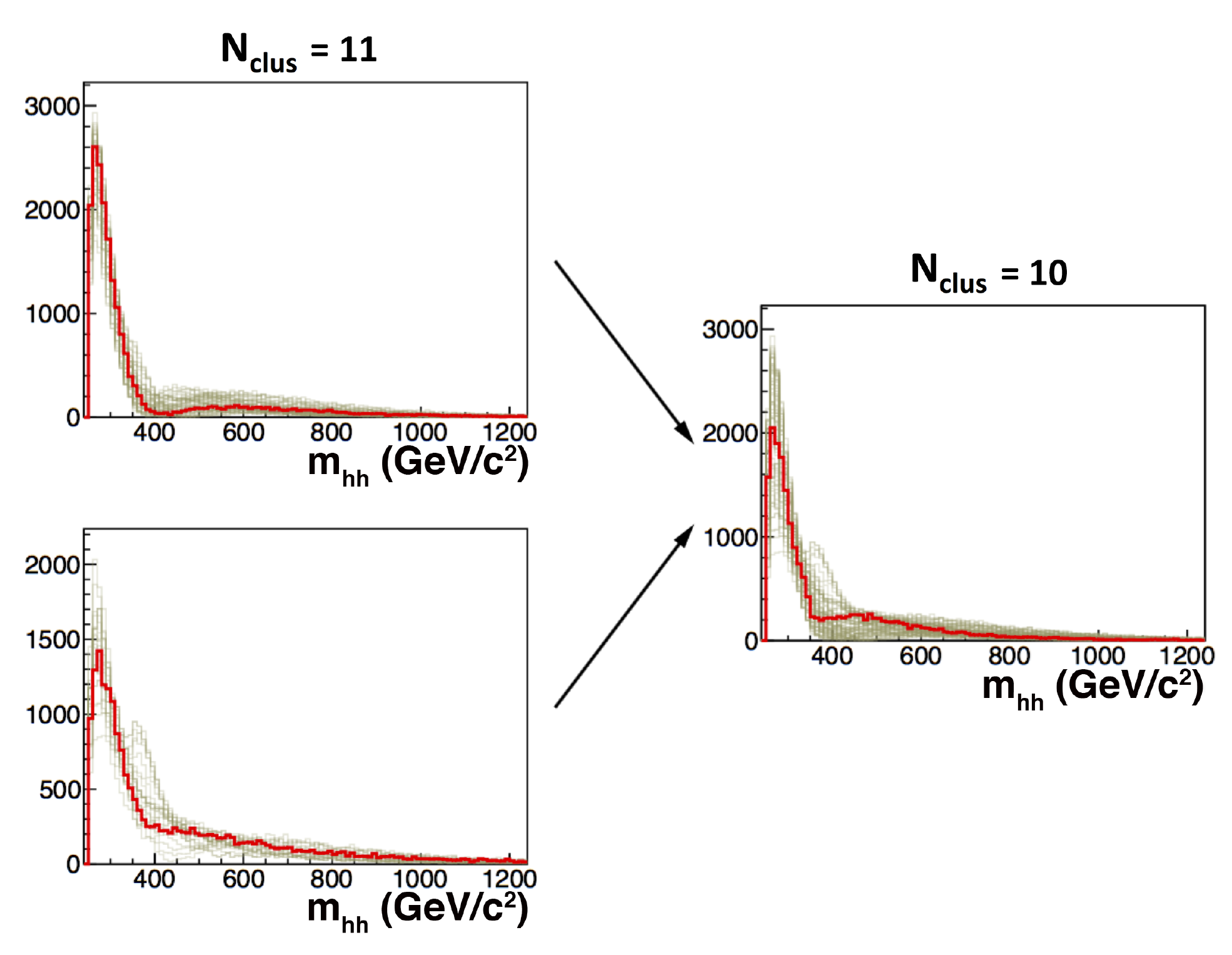} &
\includegraphics[width=0.48\textwidth, angle =0 ]{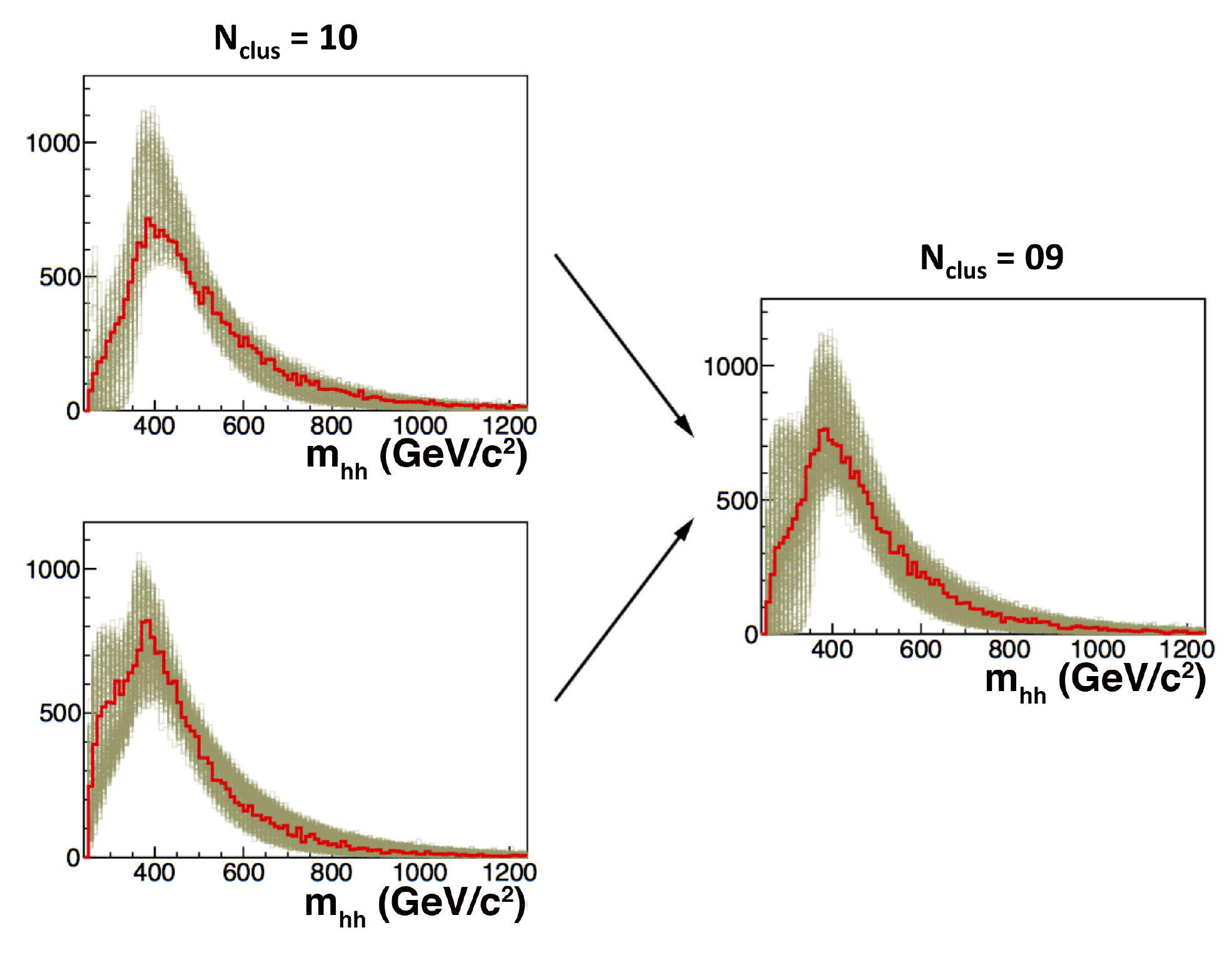} \\

\end{tabular}
\caption{\small 
Distribution of the invariant mass $m_{hh}$ of Higgs boson pairs as pairs of clusters get merged into a single one, for different values of $N_{clus}$. The red distribution is the benchmark of the cluster. The merging of clusters due to the reduction in the number of $N_{clus}$ is highlighted. It is evident that passing from $N_{clus} = 13$ to $N_{clus} = 12$ the uniformity of the distributions inside the merged cluster remains good, while subsequent mergings worsen the intra-cluster homogeneity. 
\label{fig:evol}}
\end{center}\end{figure*}

\subsection{Kinematical Sampling with $N_{clus} = 12$}

The parameter space values of the benchmarks obtained with $N_{clus} = 12$ are listed in Table~\ref{tab:bench}~\footnote {The full set of results up to $N_{clus} =20$ can be found in~\cite{fullResuts}.}. The benchmarks distribute fairly evenly in the space of model parameters, without concentrations in specific corners of phase space; furthermore, both samples with and without Higgs-gluon contact interactions are represented in the set.

Figure~\ref{fig:mlj} shows the $m_{hh}$ and $|\cos \theta^{*}|$ distributions for all the samples considered in the five-dimensional parameter space, grouped into twelve clusters by the procedure described in Sec.~\ref{sec:cluster}. Cluster 3 includes the SM point while cluster 4 includes the sample with unique contribution from the box diagram ($\kappa_{\lambda} = 0.0$,  $\kappa_t = 1.0$, and $c_2  = c_g = c_{2g} =0$). Cluster 8, which presents the characteristic doubly-peaked $m_{hh}$ distribution, includes the sample with the maximal interference between the box and triangle contributions in the SM couplings scenario, {\em i.e.} the point defined by ($\kappa_{\lambda} = 2.4$,  $\kappa_t = 1.0$, and $c_2  = c_g = c_{2g} =0$).

The clustering is clearly driven by the $m_{hh}$ variable. The impact of anomalous physics in $|\cos \theta^{*}|$ is expected to be small because all the different operators in our parametrization are predominantly s-wave (see for example~\cite{Plehn:1996wb}). This is evident in Fig.~\ref{fig:mlj}, where only few samples exhibit a non-flat structure in $|\cos \theta^{*}|$; these correspond to points of parameter space where there is a maximal interference between different terms, as in  cluster 8. All spin 2 structures (at the level of $D \leq 6$ operators, i.e. to leading approximation) come just from the box diagram. The study of the $\gamma\gamma\, b\bar{b}$ final state of {\em hh} decay is expected to be the most sensitive probe to local changes in the $m_{hh}$  spectrum; however other decay channels, such as the $WW\, b\bar{b}$ or the $b\bar{b}\, b\bar{b}$ one, could also in principle be sensitive to small shape variations in different regions of hard sub-process energy, especially when multi-variate analysis techniques are implemented. With increased statistics of the available data, fine structures in the kinematics -in particular in the $m_{hh}$ distribution, {\em e.g.} in clusters 2, 5, and 8- will become more interesting and may call for a more specific study of the corresponding regions of parameter space. 

In Fig.~\ref{fig:add} we show the distribution of the Higgs boson $p_{T}^{h}$ (which is the same for both Higgs bosons at generator level) and the longitudinal momentum of the Higgs boson with the highest energy in the laboratory frame, $|p_z^h|$. Figures~\ref{fig:mlj} and~\ref{fig:add} visually confirm that $m_{hh}$ and $|\cos \theta^{*}|$ constitute a robust choice of variables to fully describe the salient features of the $2~\to~2$ process. The features shown in Fig.~\ref{fig:add} are more directly connected with experimental selections and acceptance cuts, and to the Higgs boson reconstruction techniques. In particular, the Higgs boson transverse momentum distributions allow one to gauge how the different clusters will be affected by baseline selections in the analyses targeting the corresponding benchmarks. The $|p_z^h|$ variable is highly homogeneous within each cluster, as a result of the good properties of the clustering performed using the $m_{hh}$ variable.

It is important to point out that the clustering procedure applies no special treatment to any of the parameter space points; yet one is especially interested in the point corresponding to the Standard Model prediction. In our clustering with $N_{clus}=12$ the SM point is included in cluster 3 and it is well represented by the relative benchmark. An experimental study of the twelve benchmarks should of course be complemented by the study of the SM case; results derived for the latter are likely to be compatible with the ones for the benchmark of cluster 3.

\clearpage

\begin{figure*}[h]\begin{center}
\includegraphics[width=0.9\textwidth, angle =0 ]{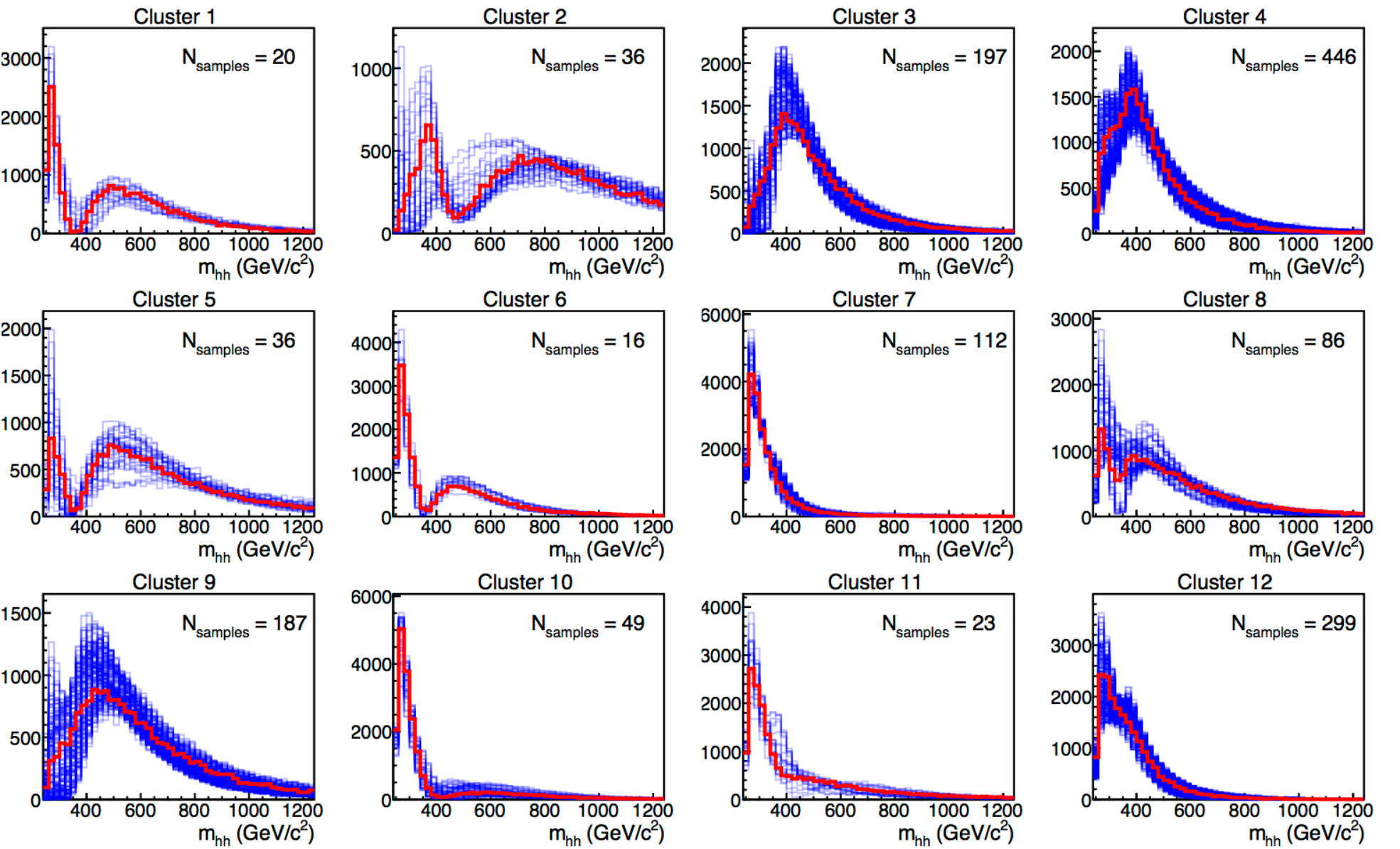}
\end{center}\end{figure*}
\vskip .2cm
\begin{figure*}[h]\begin{center}
\includegraphics[width=0.9\textwidth, angle =0 ]{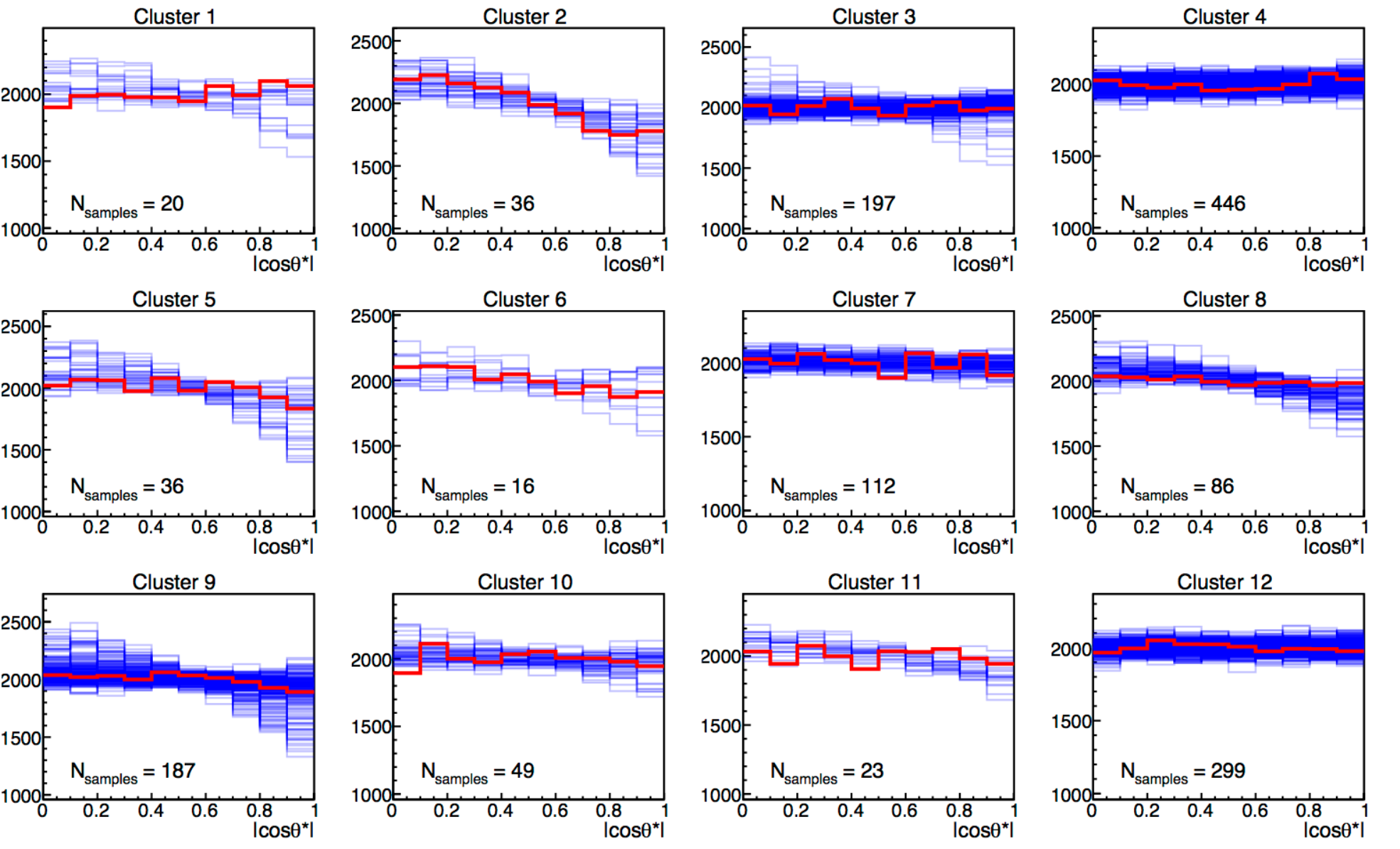}
\caption{\small 
Generation-level distributions of di-Higgs boson mass $m_{hh}$ (top three rows) and emission angle $|\cos \theta^{*}|$ (bottom three rows) for the clusters identified by the choice $N_{clus} = 12$. The red distributions correspond to the benchmark sample in each cluster, while the blue ones describe the other members of each cluster. Cluster 3 contains the SM sample.
\label{fig:mlj} }
\end{center}\end{figure*}

\clearpage

\begin{figure*}[h]\begin{center}
\includegraphics[width=0.9\textwidth, angle =0 ]{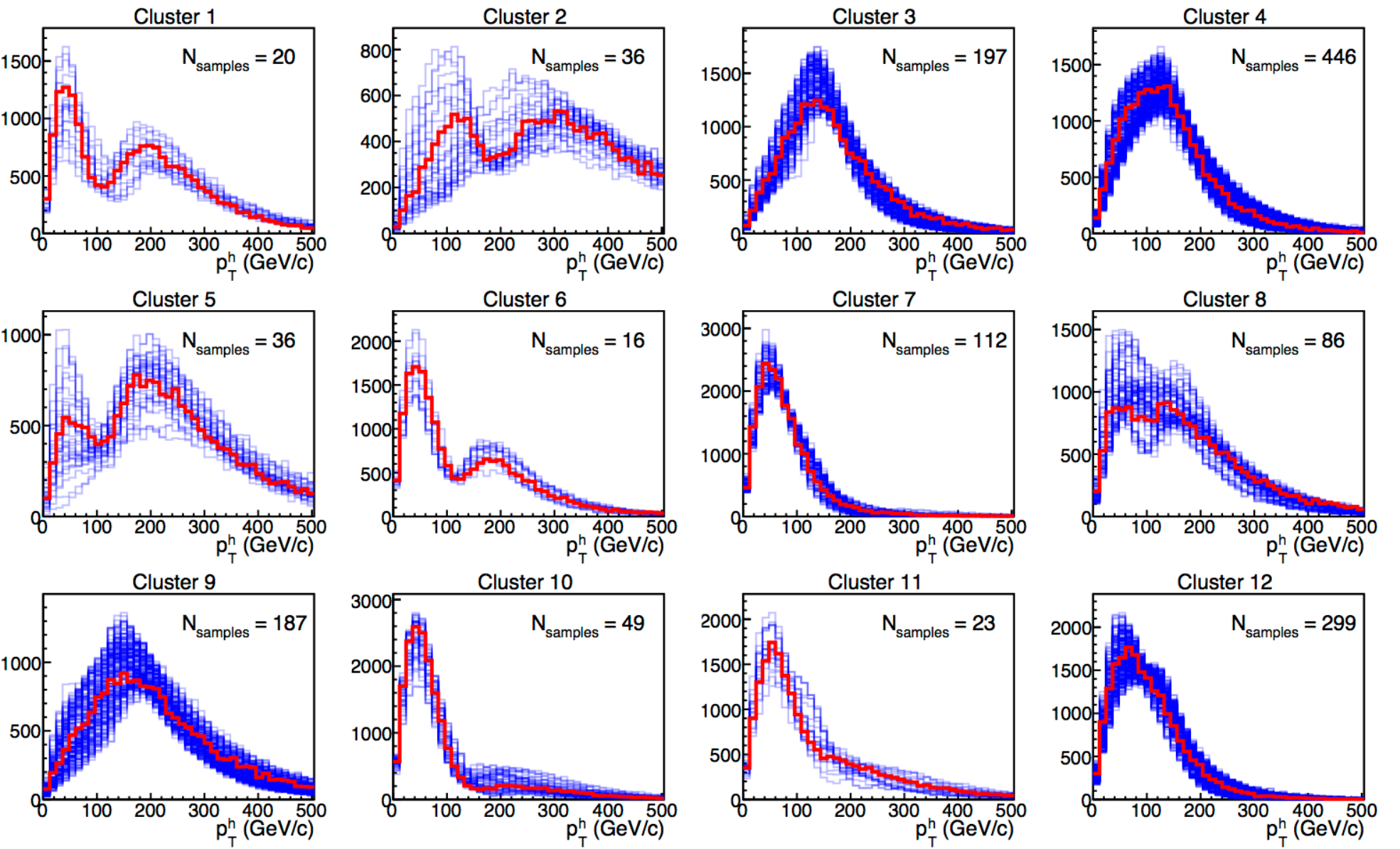}
\end{center}\end{figure*}
\vskip .2cm
\begin{figure*}[h]\begin{center}
\includegraphics[width=0.9\textwidth, angle =0 ]{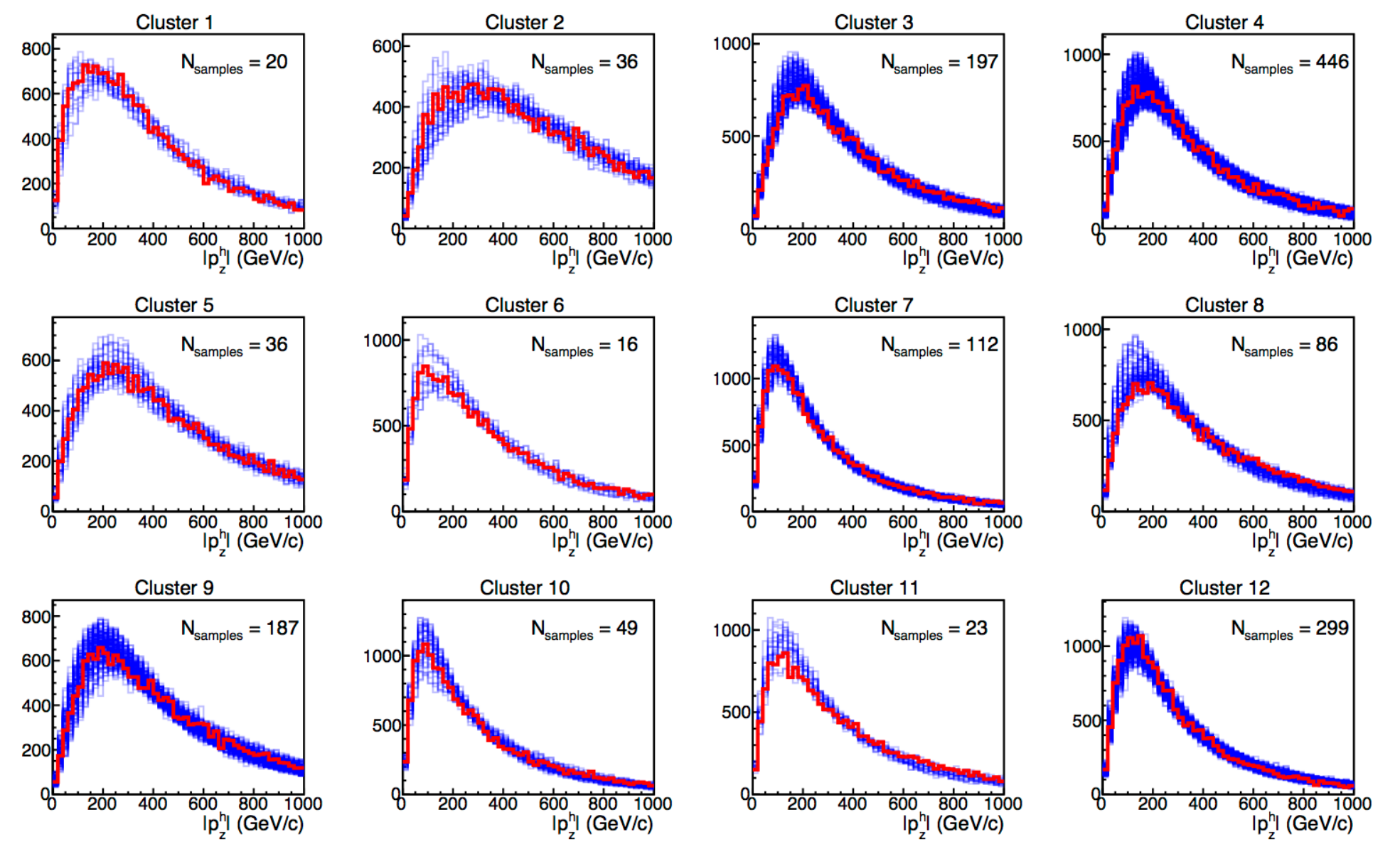}
\caption{\small 
Generation-level distributions of Higgs boson transverse momentum $p_T$ (top three rows) and absolute value of longitudinal momentum $|p_z^h|$ of the most energetic Higgs boson (bottom three rows) for the clusters identified by the choice $N_{clus}=12$. The red distributions correspond to the benchmark sample in each cluster, while the blue ones describe the other members of each cluster. Cluster 3 contains the SM sample.
\label{fig:add} }
\end{center}\end{figure*}

\clearpage

\begin{table}[h]
\centering
\small{
\begin{tabular}{rccccc}
\toprule
Benchmark & $\kappa_{\lambda}$ & $\kappa_{t}$ & $c_{2}$	& $c_{g}$ & $c_{2g}$ \\
\midrule
1 &	7.5	 & 1.0	 &	-1.0	& 0.0	& 0.0 \\
2 &	1.0	 & 1.0	 &	0.5		& -0.8	& 0.6 \\
3 &	1.0	 & 1.0	 &	-1.5	& 0.0	& -0.8 \\
4 &	-3.5 & 1.5  &	-3.0	& 0.0	& 0.0 \\
5 &	1.0	 & 1.0	 &	0.0		& 0.8	& -1 \\
6 &	2.4	 & 1.0	 &	0.0		& 0.2	& -0.2 \\
7 &	5.0	 & 1.0	 &	0.0		& 0.2	& -0.2 \\
8 &	15.0 & 1.0	 &	0.0		& -1	& 1 \\
9 &	1.0	 & 1.0	 &	1.0		& -0.6	& 0.6 \\
10 &	10.0 & 1.5   &	-1.0	& 0.0	& 0.0 \\
11 &	2.4	 & 1.0	 &	0.0		& 1		& -1 \\
12 &	15.0 & 1.0	 &	1.0		& 0.0	& 0.0 \\ \midrule 
SM &	1.0 & 1.0	 &	0.0		& 0.0	& 0.0 \\
\bottomrule
\end{tabular}
}
\caption{\small Parameter values of the twelve benchmarks and the Standard Model point.  \label{tab:bench}}
\end{table}

\clearpage
\subsection{Maps of the Clusters in the Parameter Space} \label{sec:temp}

In this section we attempt a direct mapping of the partition of the parameter space into the twelve regions found to produce homogeneous kinematical densities, using the choice of $N_{clus} = 12$. We organize our results in slices of parameter space, plotting the distribution of the clusters in each of them. 
There is no logical ordering in the numbering of the clusters; we choose markers of different shape and colour to describe how clusters spread along the different parameter space regions.  
Figure~\ref{fig:tempSM} shows the clusters distribution in the $\kappa_{t}\times\kappa_{\lambda}$ plane, which we will call SM-like plane. The iso-contours of constant cross section $\sigma_{hh}$ as computed in Sec.~\ref{sec:CX1} are shown by gray lines.

We point out how the parameter space region around the SM benchmark in the SM-like plane is especially interesting. At LO, changes in the top Yukawa parameter as small as 30\% and/or in the Higgs trilinear coupling of O(1) times the SM drive modifications of the differential cross section in $p_T^h$ from single-peaked structures to more complex two-peaked shapes where the peaks are separated by $O(100)$ GeV. As a logical corollary of what is noted above, however, one should expect the kinematical behaviour of the SM benchmark to be quite sensitive to the accuracy of the theoretical calculation. This accidental sensitivity of the kinematic behaviour to parameter values is due to the fact that the SM point is located near a cross section minimum, where there are fine cancellations between triangle and box diagrams.

\begin{figure*}[h]\begin{center}
\includegraphics[width=0.48\textwidth, angle =0 ]{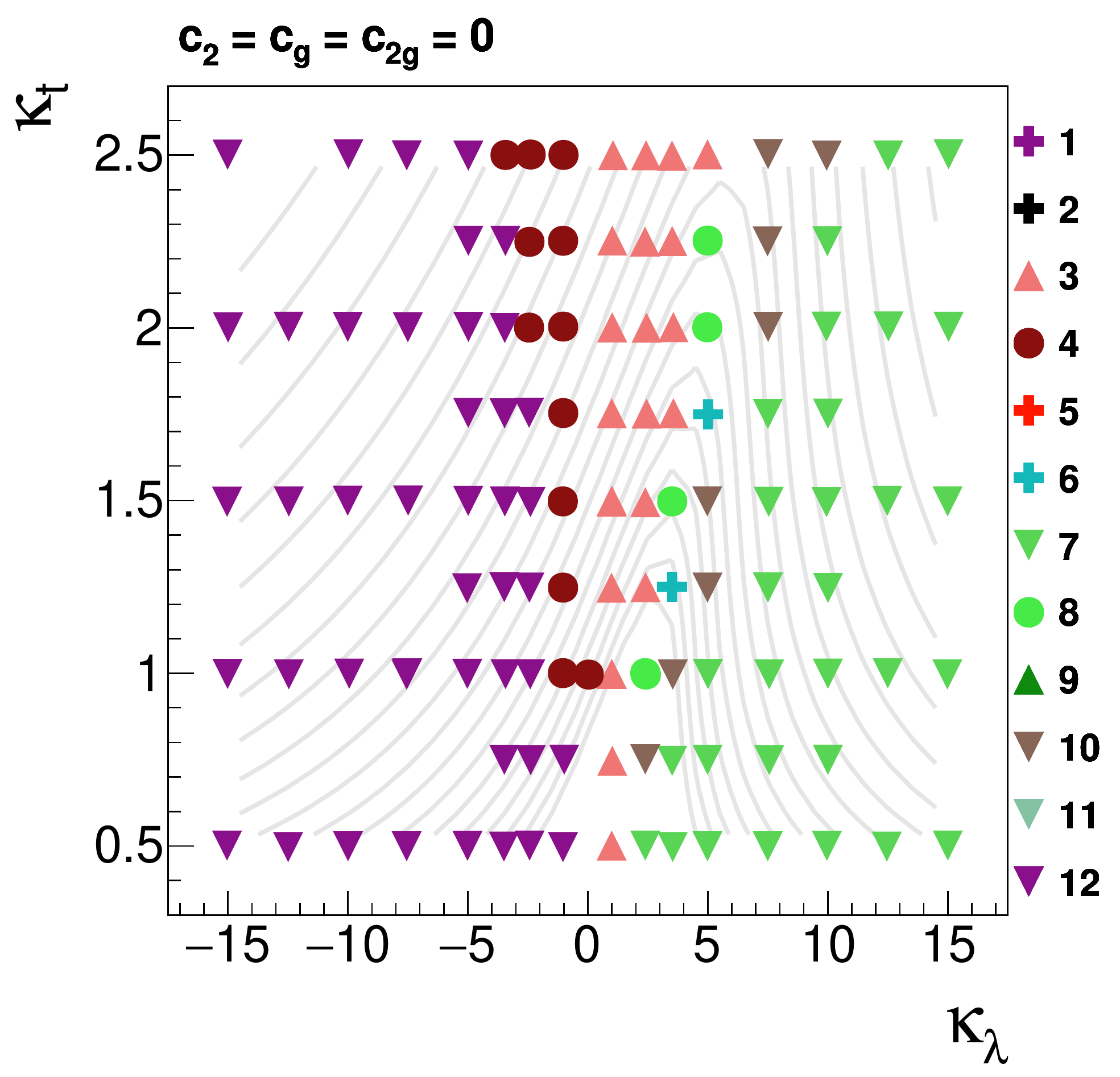}
\caption{\small
Distribution of points in the $\kappa_{\lambda} \times \kappa_t$  plane that contains the SM point. Downward-pointing triangles symbolize clusters where the benchmark has Higgs boson $p_T$ peaking at around 50 GeV or at a smaller value. Circles describe clusters whose benchmark has Higgs boson $p_T$ peaking around 100 GeV. Upward-pointing triangles describe clusters where the benchmark has Higgs boson $p_T$ peaking around 150 GeV or more.  Finally, crosses describe clusters that show a double peaking structure in the $p_T^h$ distribution. 
\label{fig:tempSM} }
\end{center}\end{figure*}

\noindent
Figure~\ref{fig:temp3D} shows the clusters in the plane $\kappa_{t}\times c_2$ for different values of $\kappa_{\lambda}$, when $c_g = c_{2g} = 0$. We observe that outside the SM-like plane there is no clear asymptotic behaviour of the event kinematics with $|\kappa_{\lambda}| \gg 1$. This confirms that asymptotic approximations of the different pieces of Eq.~\ref{eq:ME} are not useful for a deep parameter space scan. Figure~\ref{fig:5Dgridcg} shows the  map of clusters in various slices of the five-dimensional parameters space, the same used in Sec.~\ref{sec:CX1} for the calculation of the cross section modifications.  
 
\begin{figure*}[h]\begin{center}
\includegraphics[width=1.0\textwidth, angle =0 ]{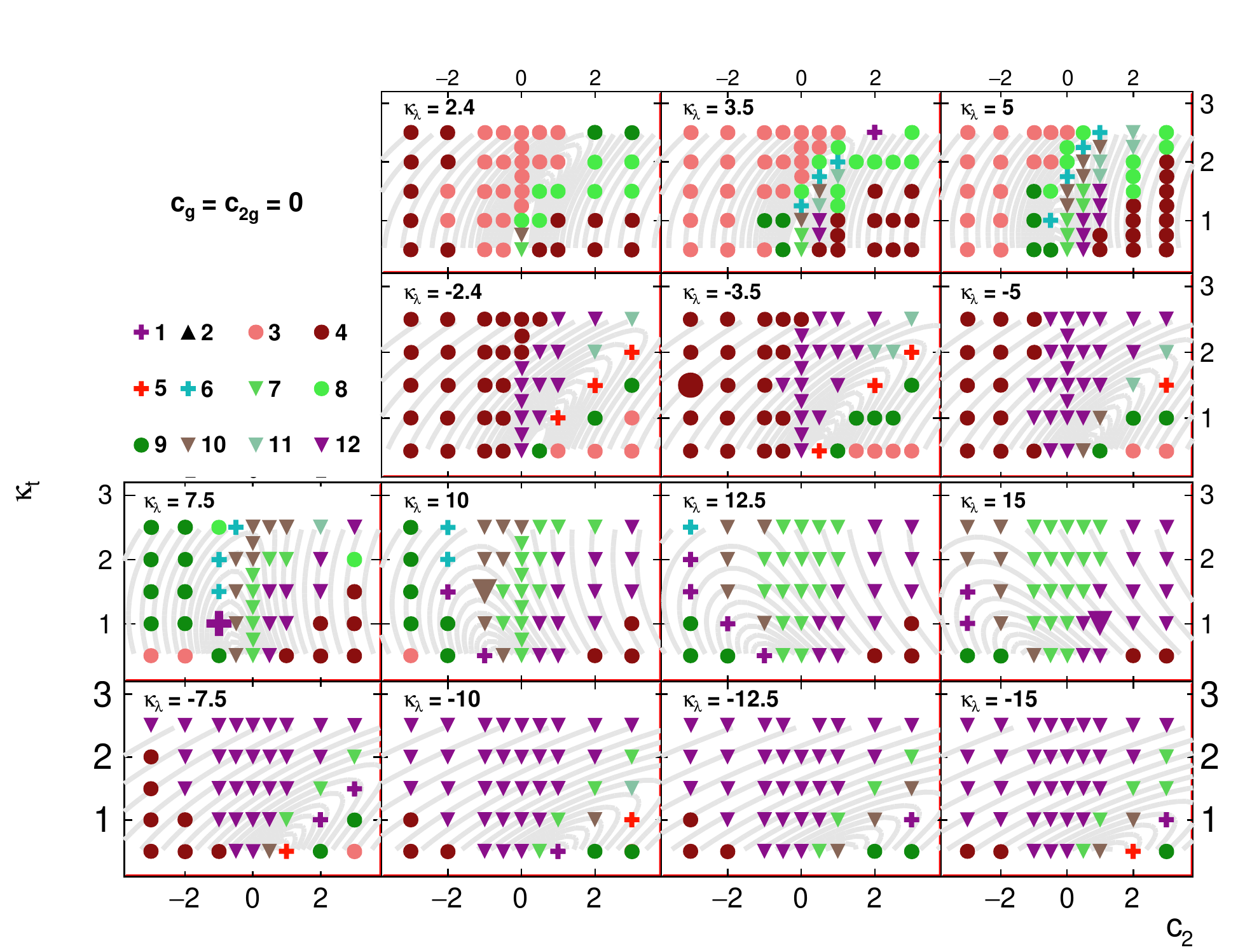}
\caption{\small
Distribution of points in the $c_2 \times \kappa_t$ plane for different values of $\kappa_{\lambda}$ when $(c_g,\, c_{2g}) = (0,0)$.
The different markers represent different regimes of Higgs boson $p_T$,  as described in the caption of Fig.~\ref{fig:tempSM}. Larger markers indicate benchmark points.
\label{fig:temp3D} }
\end{center}\end{figure*}

\begin{figure*}[h]\begin{center}
\includegraphics[width=0.3\textwidth, angle =0 ]{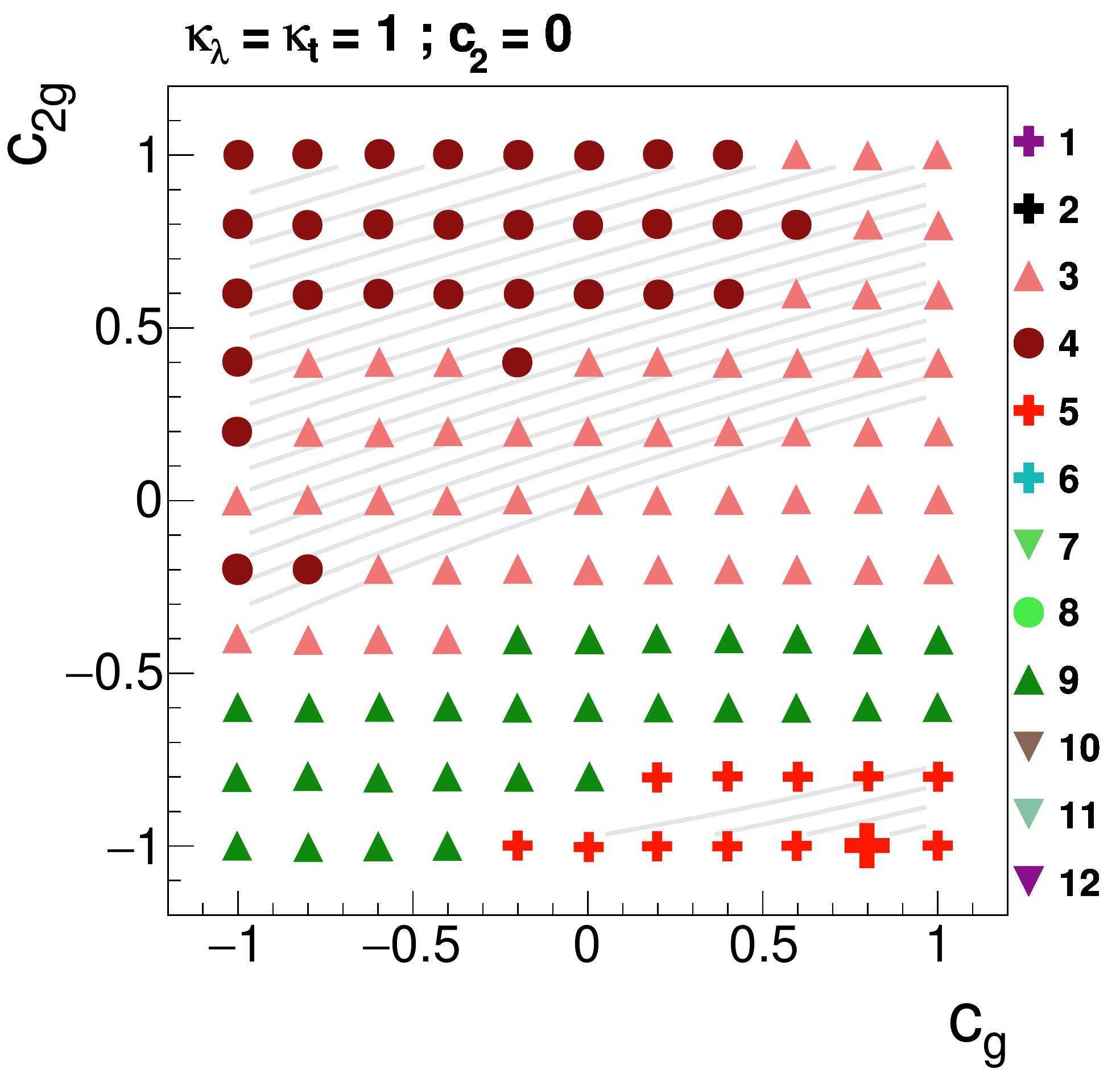}
\includegraphics[width=0.3\textwidth, angle =0 ]{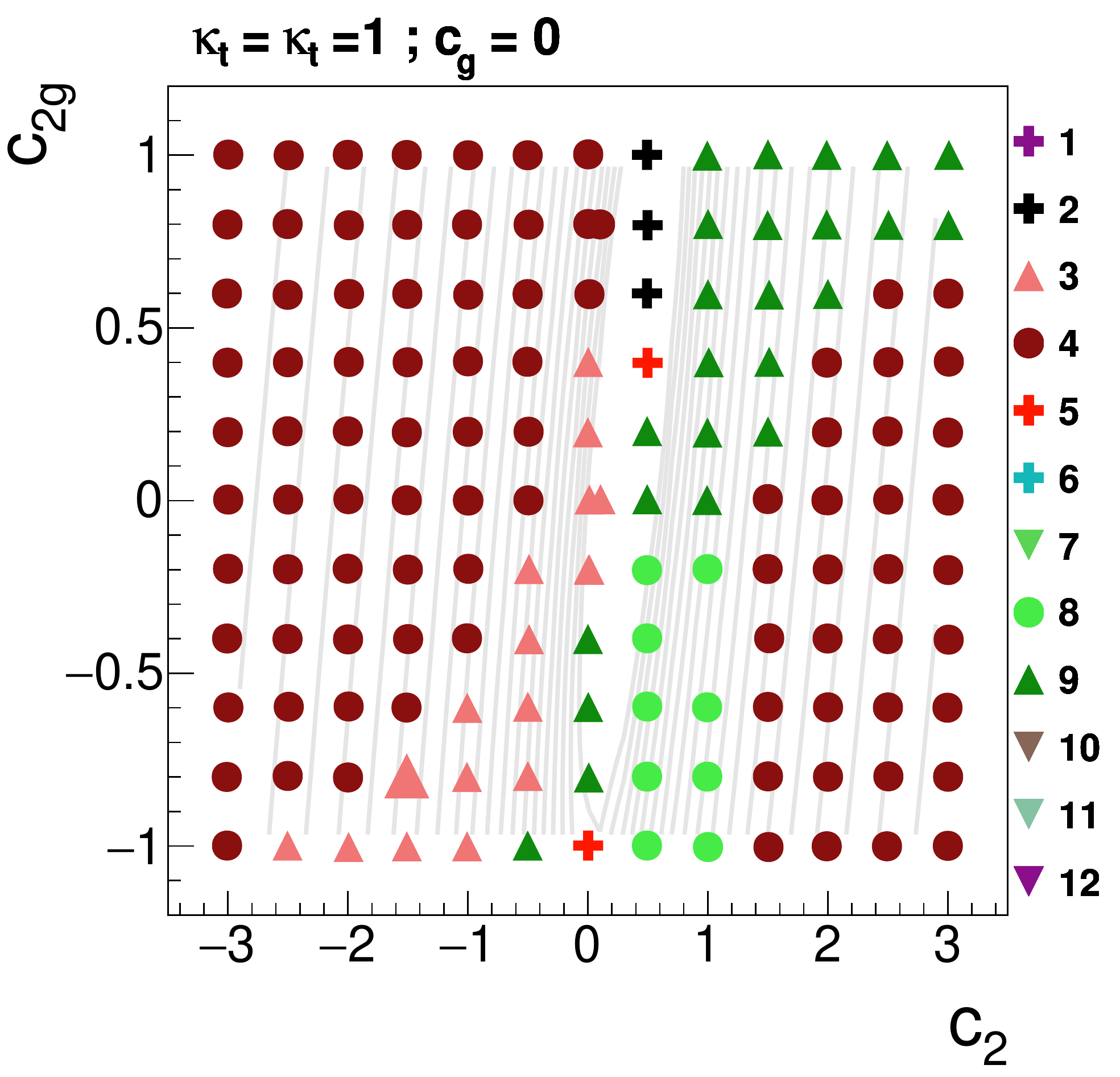}
\includegraphics[width=0.3\textwidth, angle =0 ]{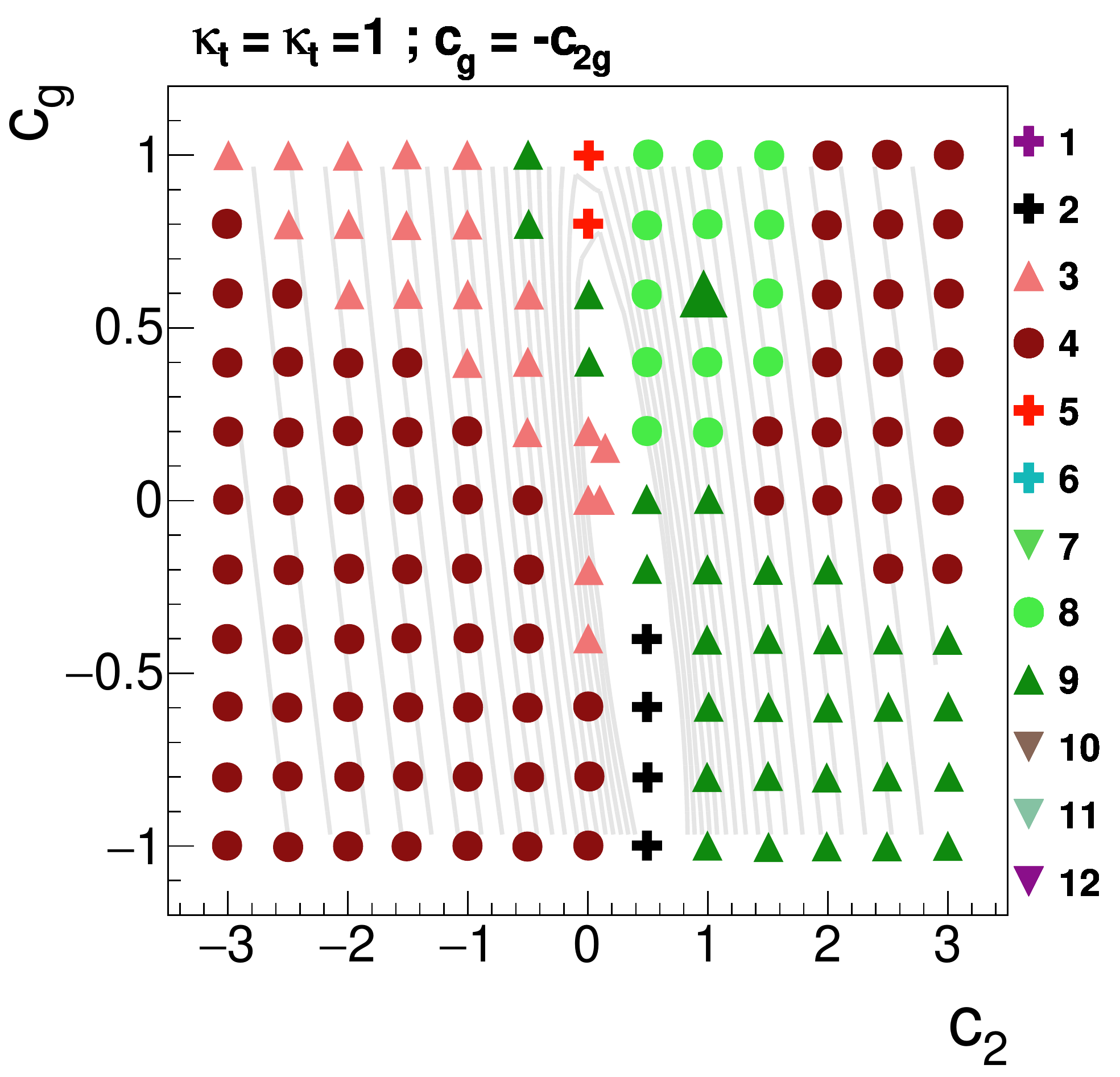}
\includegraphics[width=0.3\textwidth, angle =0 ]{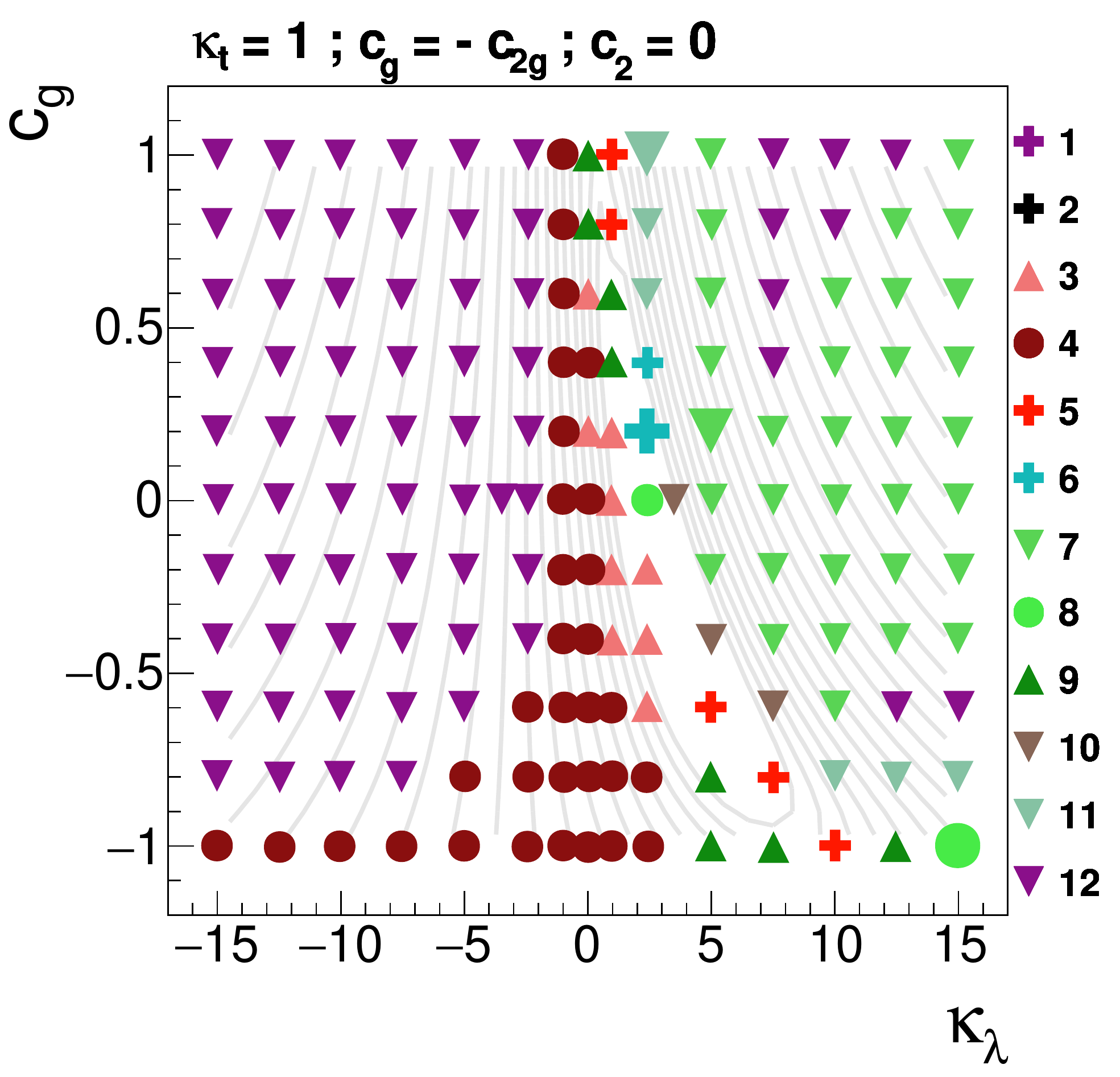}
\includegraphics[width=0.3\textwidth, angle =0 ]{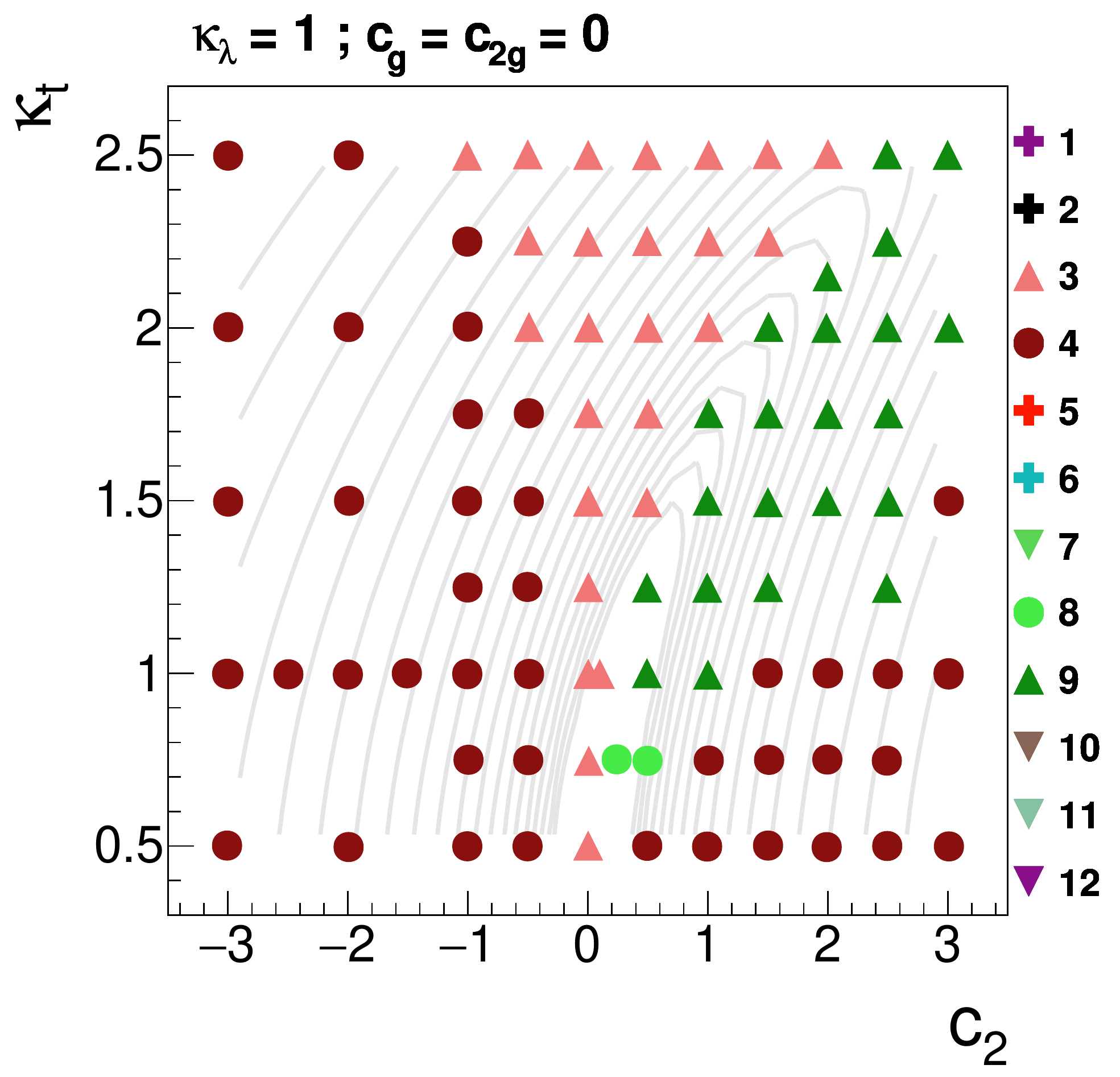}
\includegraphics[width=0.3\textwidth, angle =0 ]{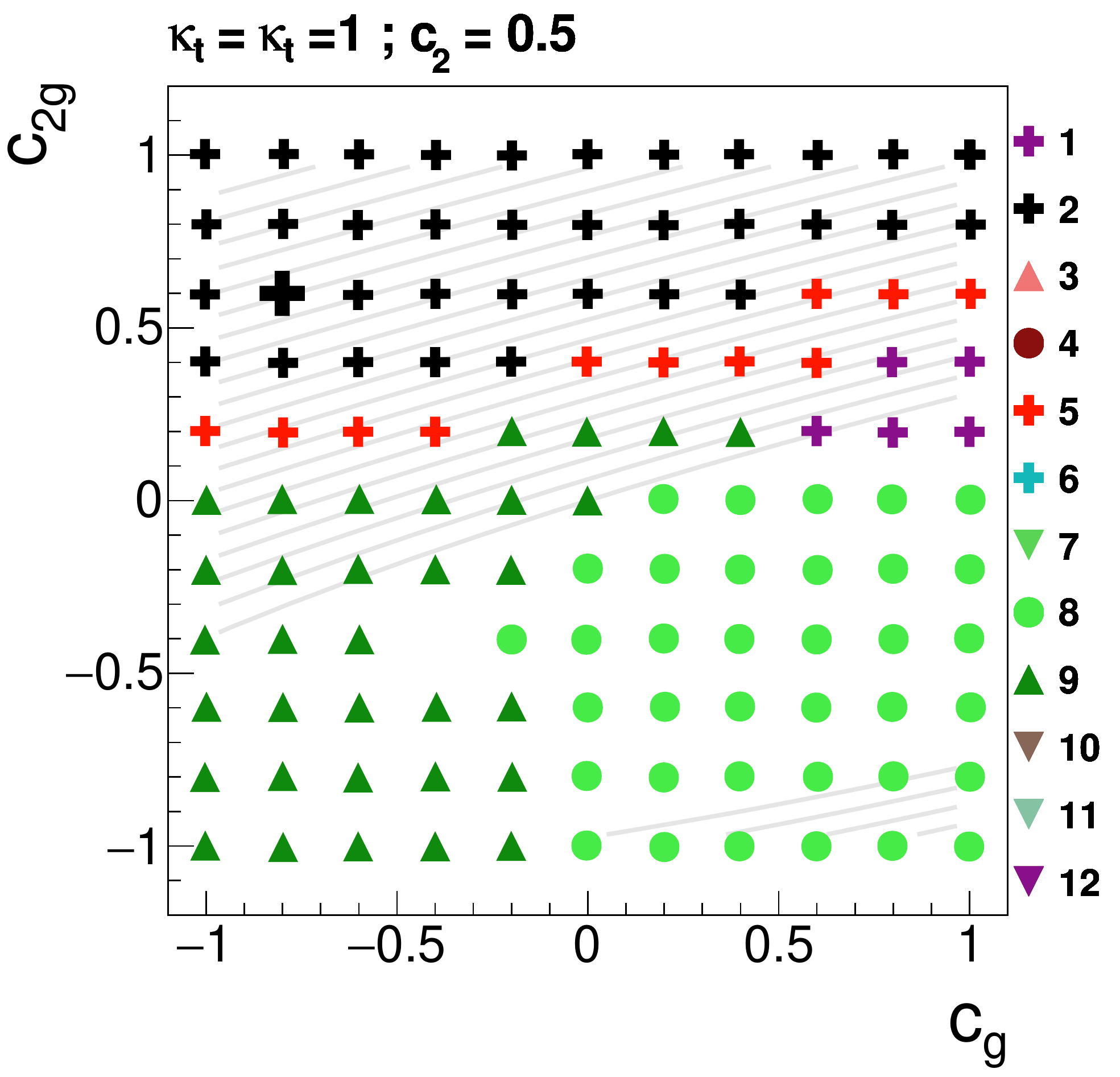}
\caption{\small
Distribution of clusters in various slices of the five-dimensional parameters space. The different shapes of the markers represent different regimes of Higgs boson $p_T$,  as described in the caption of Fig.~\ref{fig:tempSM}. Larger markers indicate benchmark points.
\label{fig:5Dgridcg}}
\end{center}\end{figure*}

\noindent
Figures~\ref{fig:tempSM} and~\ref{fig:temp3D} suggest that the largest modifications in the final state kinematics are tightly related to the minima of the cross section. Since all the matrix-element pieces not related with interferences ($|M_i|^2$) are positive-definite, the minima of the cross section in each slice of parameter space are partially a reflection of regions where the interferences between the different processes are large in comparison with the other ME pieces. The correspondence however is not bilateral: the balance between the non-interference terms can also drive visible changes in shapes while not affecting much the total cross section. 

As an additional qualitative proof of the close correspondence between the cross section minima and the regions of largest variation of the density of the kinematical distributions in the final state, we show in Fig.~\ref{fig:TSandCX} a few maps of the cross section of di-Higgs boson production in two-dimensional subspaces of the five model parameters, with overlaid colour maps describing the magnitude of the point-to-point variations in the value of the log-likelihood test statistic defined in Sec.~\ref{sec:cluster}. The latter describe the speed with which the $m_{hh}$ and $|\cos \theta^{*}|$ distributions vary, highlighting the effect of the cancellation of diagram contributions mentioned above.

\begin{figure*}[h]\begin{center}
\includegraphics[width=1.\textwidth, angle =0 ]{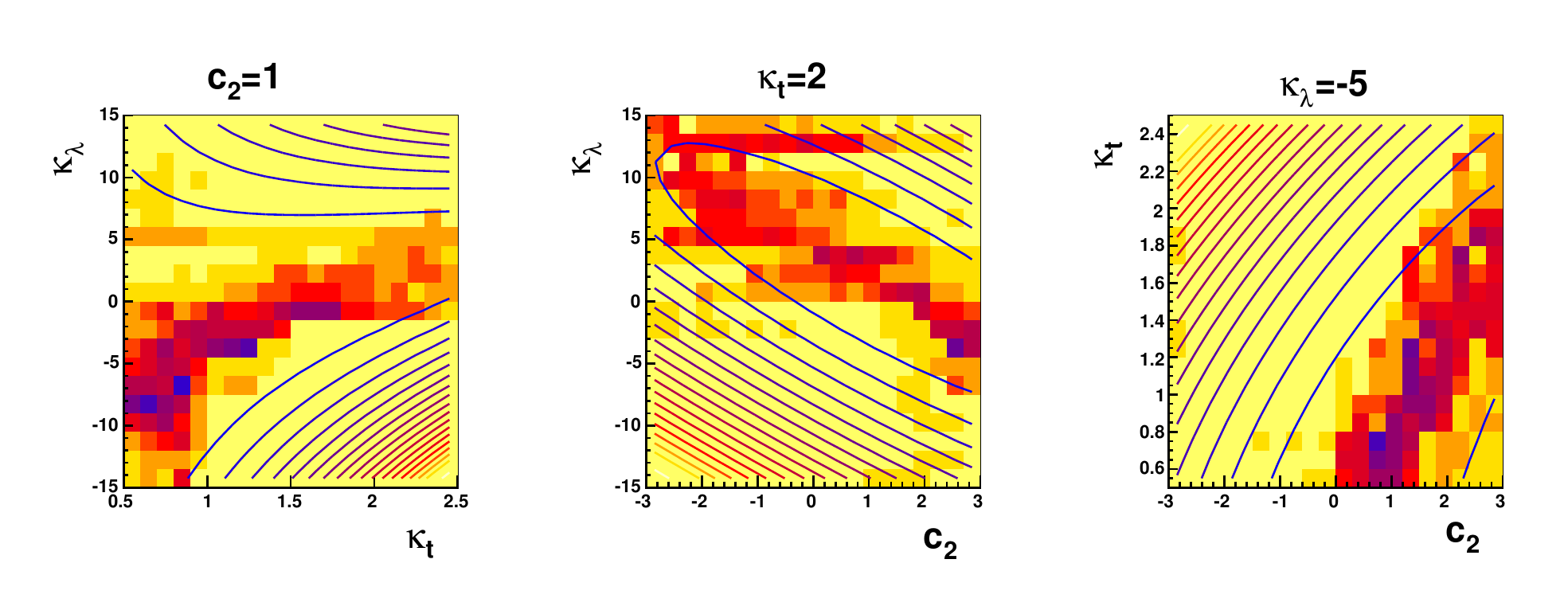} \caption{\small
Superposition of isolines of cross section and colour maps of the speed at which the likelihood test statistic described in Sec.~\ref{sec:cluster} varies as one moves around in three selected two-dimensional surfaces of the five-dimensional parameter space of BSM theories. The cross section decreases in the direction where the density of isolines decreases.  Blue and red tones in the colour maps indicate the highest variation in the $TS$ values; the colour scale is arbitrary. The behaviour observed in the graphs is common to all investigated two-dimensional planes.
\label{fig:TSandCX} }
\end{center}\end{figure*}

\section{Conclusions}
\label{sec:conclu}
\vskip .3cm

The study of Higgs pair production processes at the LHC may evidence the existence of BSM physics in the form of anomalous (self-)couplings of the Higgs boson. While both the total Higgs pair-production cross section and the kinematics of the final state depend on those couplings, it is the latter that impact the most the design of experimental techniques aimed at measuring the process. In this article we described a procedure to define suitable benchmark points in the multi-dimensional parameter space spanning the possible value of the anomalous couplings. The procedure optimally chooses the benchmarks such that the study of the resulting physics has the largest impact on the exploration of the parameter space. 

The technique we propose is based on the definition of a test statistic measuring the similarity of the kinematics of the Higgs boson pairs in the final state resulting from different parameter space points, and a suitable clustering procedure to group parameter space points together. Although it finds a very profitable application to the case of Higgs boson pair production at the LHC, the technique is quite general and may successfully be employed in other physics studies.

We study gluon-fusion-initiated di-Higgs boson production in 13 TeV proton-proton collisions and examine an extensive but not exhaustive set of subspaces of the five-dimensional space of anomalous couplings. We find that twelve benchmarks are sufficient to describe to a reasonable level of approximation the possible different kinematic densities that may arise from arbitrary combinations of the parameters. We argue that an experimental study which focuses on those twelve scenarios should maximize the impact on the exploration of the parameter space, without leaving unexplored ``holes''.

The grouping of parameter space points is also meant to allow one to extrapolate the results of an experimental search performed on a benchmark point to all other points of the cluster which contains the benchmark. Whether such an appealing plan is feasible remains to be proven, as it depends on the homogeneity of the intra-cluster kinematics as well as on the statistical power of the experimental data. A detailed study of the degree of validity of such extrapolations will be the subject of future investigations.

\clearpage
\appendix
\section{Cross Section Coefficients}
\label{sec:app}

In this appendix we summarize the procedure we followed to fit the coefficients $A_i$ in the cross section ratio, Eq.~\ref{eq:cx}. In principle to fix the fifteen coefficients in a recursive way one needs to calculate the total cross section only for fifteen selected  points in parameter space. The cross section estimates are however obtained with a Monte Carlo generator and they may contain intrinsic errors, related for example to the finite statistics in phase space integration. Moreover, the final result also includes uncertainties due to PDF errors and missing higher orders, captured by scale variations.  

In order to properly account for the effects mentioned above, and in particular to judge the associated stability of the fitted coefficients of Eq.~\ref{eq:cx}, the fit must be performed using a large data sample. Such a study is described in~\cite{CarvalhoAntunesDeOliveira:2130724}, and the result is reported in Table~\ref{tab:coef}. 

In addition to the coefficients for the cross sections at 13 TeV, for completeness we also provide the values of the cross section coefficients for $pp$ collisions at 7, 8, 14 and 100 TeV centre-of-mass energy. 
The theory uncertainties on the cross sections defined by Eq.~\ref{eq:cx} and Table~\ref{tab:coef} are also evaluated and can be found in~\cite{CarvalhoAntunesDeOliveira:2130724}. 

\begin{table}[h]
\centering
\footnotesize{
\begin{tabular}{rccccc}
\toprule
$\sqrt{s} $ & 7 TeV & 8 TeV & 13 TeV & 14 TeV & 100 TeV\\
\midrule
$A_1$  &  2.21  &  2.18  &  2.09  &  2.08  &  1.90  \\
$A_2$  &  9.82  &  9.88  &  10.15  &  10.20  &  11.57  \\
$A_3$  &  0.33  &  0.32  &  0.28  &  0.28  &  0.21  \\
$A_4$  &  0.12  &  0.12  &  0.10  &  0.10  &  0.07  \\
$A_5$  &  1.14  &  1.17  &  1.33  &  1.37  &  3.28  \\
$A_6$  &  -8.77  &  -8.70  &  -8.51  &  -8.49  &  -8.23  \\
$A_7$  &  -1.54  &  -1.50  &  -1.37  &  -1.36  &  -1.11  \\
$A_8$  &  3.09  &  3.02  &  2.83  &  2.80  &  2.43  \\
$A_9$  &  1.65  &  1.60  &  1.46  &  1.44  &  3.65  \\
$A_{10}$  &  -5.15  &  -5.09  &  -4.92  &  -4.90  &  -1.65  \\
$A_{11}$  &  -0.79  &  -0.76  &  -0.68  &  -0.66  &  -0.50  \\
$A_{12}$  &  2.13  &  2.06  &  1.86  &  1.84  &  1.30  \\
$A_{13}$  &  0.39  &  0.37  &  0.32  &  0.32  &  0.23  \\
$A_{14}$  &  -0.95  &  -0.92  &  -0.84  &  -0.83  &  -0.66  \\
$A_{15}$  &  -0.62  &  -0.60  &  -0.57  &  -0.56  &  -0.53  \\
\bottomrule
\end{tabular}
}
\caption{\small Coefficients of our fit to the  cross section modifications of double Higgs production  in proton-proton collisions with respect to the SM benchmark (Eq.~\ref{eq:cx}).  See~\cite{CarvalhoAntunesDeOliveira:2130724} for the relative theory uncertainties. 
 \label{tab:coef}}
\end{table}

\clearpage
\acknowledgments
We wish to thank Maxime Gouzevich, Olivier Bondu, Fabio Maltoni, Eleni Vryonidou, Benoit Hespel, Andreas Papaefstathiou and Jose Zurita for useful discussions. A.O would like to extend this list to Andrea Wulzer, Giuliano Panico and Christoph Englert.  We also thank  Amina Zghiche, Debdeep Ghosal and Serguei Ganjour for further cross checks.  A.C. and F.G. would like to express special thanks to the Mainz Institute for Theoretical Physics (MITP) for its hospitality. A.C. is  supported by  MIURFIRB RBFR12H1MW grant. The research of F.G is supported by a Marie Curie Intra European Fellowship within the 7th European Community Framework Programme (grant no. PIEF-GA-2013-628224).

\printbibliography

\end{document}